\documentclass[sn-mathphys,Numbered]{sn-jnl}

\usepackage{graphicx}%
\usepackage{multirow}%
\usepackage{amsmath,amssymb,amsfonts}%
\usepackage{amsthm}%
\usepackage{mathrsfs}%
\usepackage{xcolor}%
\usepackage{textcomp}%
\usepackage{manyfoot}%
\usepackage{booktabs}%
\usepackage{algorithm}%
\usepackage{algorithmicx}
\usepackage{geometry}
\usepackage{url}

\unnumbered 

\begin{document}

\title[Article Title]{Unpacking Polarization: Antagonism and Alignment in Signed Networks of Online Interaction}


\author*[1]{\fnm{Emma} \sur{Fraxanet}}\email{emma.fraxanet@upf.edu}

\author[3]{\fnm{Max} \sur{Pellert}}

\author[4]{\fnm{Simon} \sur{Schweighofer}}

\author[1]{\fnm{Vicenç} \sur{Gómez}}

\author[2,5]{\fnm{David} \sur{Garcia}}

\affil*[1]{\orgdiv{Department of Information and Communication Technologies}, \orgname{Pompeu Fabra University}, \orgaddress{\street{Tànger 122-140}, \city{Barcelona}, \postcode{08018}, \country{Spain}}}

\affil[2]{\orgname{Complexity Science Hub}, \orgaddress{\street{Josefst\"after Strasse 39}, \city{Vienna}, \postcode{1080}, \country{Austria}}}

\affil[3]{\orgdiv{Chair for Data Science in the Economic and Social Sciences}, \orgname{University of Mannheim}, \orgaddress{\street{L15, 1–6}, \city{Mannheim}, \postcode{68161}, \country{Germany}}}

\affil[4]{\orgdiv{Department of Media \& Communication}, \orgname{Xi'an Jiaotong-Liverpool University}, \orgaddress{\street{Suzhou Industrial Park}, \city{Suzhou}, \postcode{215123}, \country{P.R.China}}}

\affil[5]{\orgdiv{Department of Politics and Public Administration}, \orgname{University of Konstanz}, \orgaddress{\street{Universit\"atstrasse 10}, \city{Konstanz}, \postcode{78464}, \country{Germany}}}


\abstract{
Political conflict is an essential element of democratic systems, but can also threaten their existence if it becomes too intense. This happens particularly when most political issues become aligned along the same major fault line, splitting society into two antagonistic camps. In the 20th century, major fault lines were formed by structural conflicts, like owners vs workers, center vs periphery, etc. But these classical cleavages have since lost their explanatory power. Instead of theorizing new cleavages, we present the FAULTANA (FAULT-line Alignment Network Analysis) pipeline, a computational method to uncover major fault lines in data of signed online interactions. Our method makes it possible to quantify the degree of antagonism prevalent in different online debates, as well as how aligned each debate is to the major fault line. This makes it possible to identify the wedge issues driving polarization, characterized by both intense antagonism and alignment. We apply our approach to large-scale data sets of Birdwatch, a US-based Twitter fact-checking community and the discussion forums of DerStandard, an Austrian online newspaper.
We find that both online communities are divided into two large groups and that their separation follows political identities and topics. In addition, for DerStandard, we pinpoint issues that reinforce societal fault lines and thus drive polarization. We also identify issues that trigger online conflict without strictly aligning with those dividing lines (e.g. COVID-19).
Our methods allow us to construct a time-resolved picture of affective polarization that shows the separate contributions of cohesiveness and divisiveness to the dynamics of alignment during contentious elections and events.
}

\keywords{Polarization, Political cleavage structures, Signed Networks, Social Media, Alignment, Antagonism }

\newgeometry{top=1in,bottom=2in,right=1.5in,left=1.5in}

\maketitle

\section{Introduction}

It is nowadays difficult to watch a news broadcast, listen to a campaign speech, or read a political commentary without coming across the term \emph{polarization}. 
It seems that, when political commentators need a catchy, one-word description of the current state of political affairs, they habitually default to \emph{polarized}. 
But this inflationary usage of the concept of political polarization lumps together very different forms of political conflict.
In a world where even apparently apolitical questions of lifestyle and taste have become associated with ideological positions \cite{dellaposta2015liberals}, it may seem like every political conflict is being fought along the lines of left versus right, neatly splitting the political spectrum into two opposed factions. 
But neither in theory nor in practice is this the only way in which political antagonism can manifest in democratic societies.

The conflation of concepts when talking about polarization also explains the seemingly ambivalent role of political antagonism in democratic societies: 
On the one hand, polarization is usually conceptualized as detrimental to political stability and efficient governance. 
On the other hand, conflict and competition are recognized as essential parts of a functioning political system.
This apparent contradiction is easily resolved by stipulating that political antagonism is not automatically detrimental to the stability of the system, as long as it is not exclusively located along the same dividing line, or \emph{cleavage}.
If political antagonism is located along multiple \emph{cross-cutting cleavages} \cite{rokkan1967geography,blau1984crosscutting}, it can actually increase systemic cohesion by putting political actors into ever-changing configurations of alliances.
In such a system, the opponents of yesterday may become the allies of tomorrow (and vice versa), which creates an incentive to maintain a minimum of civility \cite{mason2016cross}. 
In contrast, if conflicts are predominantly organized along a single cleavage, political actors will always find themselves alongside, and across from, the same group of people. It is easy to see why in such a system civility tends to be replaced by partisan hostility and political sectarianism \cite{finkel2020political}. 

The analysis of cleavage structure has been a central concern for political scientists (especially in Europe) since the seminal work of Lipset and Rokkan in 1967 \cite{lipset1967party}. They theorized that party systems in Western democracies are the results of four basic societal conflicts: center vs. periphery, state vs. church, owner vs. worker, and land vs. industry, which are present to differing degrees in different societies.  
The four cleavages initially introduced by Lipset and Rokkan in the 1960s have since lost a large degree of their explanatory power \cite{franklin1992decline}.
New cleavages have been proposed by various authors, determined, for example, by conflicts around globalization \cite{kriesi2008west}, migration \cite{ford2020changing}, or European integration \cite{hooghe2018cleavage}.
However, it has been criticized that, similar to 'polarization', the term 'cleavage' has been overexpanded, and thus lost most of its meaning, serving now merely as a redescription of differences in political attitudes among the electorate \cite{bartolini2007identity,goldberg2020evolution}.

In this study, we identify and analyze two distinct factors of political polarization: 
First, the degree of \emph{Antagonism} in a community, a metric reflecting the prevalence of negativity in the interactions that are triggered by a controversial issue.
And second, the degree of \emph{Alignment} of a community around an issue, reflecting how much the issue 'fits', and thereby reinforces, the main dividing lines in a community.
Political polarization can then be defined as the product of Antagonism and Alignment, both of which have to be present for a political system to fission into radically opposed factions.  

Our quantification of Antagonism and Alignment is based on the identification of cohesive groups in networks of signed relations as well as the cleavages separating them. In such signed networks, each node represents an individual and their relations are represented by positive or negative edges. 

In social media, positive interactions are captured by liking, praising, forming friendships, or establishing trust, while negative interactions are captured by disliking, toxic behavior, hostility, or distrust. By considering explicitly negative interactions within social media, we gain a deeper understanding of community structures and relations than by only analyzing positive interactions. For example, relying only on positive interaction data creates biases that lead to  an overestimation of online fragmentation and distorted pictures of the polarization of a community \cite{guerra2013measure} \cite{keuchenius2021important}. This is particularly important when assessing the degree of political polarization in social media use, which might have been overstated due to missing information on negative interactions \cite{barbera2015tweeting}.

Balance theory \cite{heiderPsychologyInterpersonalRelations1958,cartwright1956structural} postulates that positive interactions happen with a higher likelihood between individuals belonging to a same group (e.g. political faction), whereas negative interactions happen predominantly between opposed factions. Balance can also be defined by the absence of cycles containing an odd number of negative edges \cite{cartwright1956structural}. In practice, real-world signed networks are not completely balanced and different definitions of partial balance have been introduced, e.g., signed triangle count\cite{leskovec2010signed}, walk-based partial balance measures \cite{estrada2014walk} or frustration-based measures \cite{aref2018measuring}. 
The latter provides a network partitioning algorithm according to a maximization of balance.

Building on balance maximizing partitions, we designed FAULTANA, our proposed pipeline for assessing Alignment and Antagonism in online interactions. Our framework can track changes in Alignment over time and compare how group structure manifests across issues in society. This way, we can discover cleavages based on high-resolution and contextualized data as a supplementary approach to theorizing specific cleavages \emph{ab initio}. Furthermore, we analyze the two independent mechanisms that contribute to Alignment, namely \textit{Cohesiveness} and \textit{Divisiveness}~\cite{aref2020multilevel}, which account for in-group agreement versus out-group disagreement. At present, out-group disaffection is the most relevant variable in the steep increase of political sectarianism, especially in the US \cite{finkel2020political}. Hence, a proper consideration of negative interactions and relations is crucial to the analysis of polarization within online systems. 

We apply this framework to two unique datasets that contain positive and negative interactions between users extracted from two different platforms: Birdwatch, the US pilot stage of a crowd-sourced fact-checking Twitter system; and DerStandard, an Austrian online newspaper with discussions on news pieces. 

The signed network data of Birdwatch and DerStandard offer a unique opportunity to directly measure positive and negative relationships, as previous research struggled to infer negative relationship information from unsigned data \cite{andres2022reconstructing,garcia2013measuring,neal2014backbone}. This difficulty is particularly pronounced in online social systems, where distinguishing between users not interacting due to animosity versus chance becomes infeasible \cite{guerra2013measure}. Even the inference of positive interactions from endorsing actions, such as retweets, has been called into question \cite{tufekci2014big}.
Some exceptions in very precise contexts exist, such as signed graphs of political elite interactions (e.g. international relations \cite{doreian2015structural,maoz2007enemy,diaz2023network,estrada2019rethinking} and the US House of Representatives \cite{aref2021identifying}), online platforms with particular functions away from general discussion (e.g. Epinions \cite{guha2004propagation}, Slashdot \cite{kunegis2009slashdot} or Wikipedia \cite{west2014exploiting,maniuBuildingSignedNetwork2011}), and inferred signed interactions from text data in Reddit \cite{pougue2021debagreement,pougue2022learning}.

Our two datasets provide information on general discussions with strong political content and explicit signed interactions of the form of positive and negative ratings based on spontaneous behavior. Both datasets also have temporal information and contextualization features encoded in news tags in the case of DerStandard and text in both datasets. While Birdwatch has been studied in previous research \cite{prollochs2022community,saeed2022crowdsourced,drolsbach2023believability,wojcik2022birdwatch,drolsbach2022diffusion,allen2022birds}, our DerStandard dataset is novel and comprises eight years of signed interaction information between regular users of news discussions.

\section{Data and Methods}\label{Methods}
\subsection{Data description and preparation}\label{Data}
We use these key features of our two data sources: (a) DerStandard: positive (+) and negative (-) ratings on postings in the forum below articles on the online newspaper page. (b) Birdwatch: agreement and disagreement between raters and their notes, which we treat as positive and negative interactions. In both cases we also have temporal information (timestamp of postings or note).

We differentiate between: (i) \textit{Interactions}: directed pairwise interactions based on the reaction of a user (rater) to the content posted by another user (author), with the timestamp corresponding to the posting of that piece of content, and (ii) \textit{Edges}: undirected and signed relations between users of the platform, based on aggregated interactions exchanged between them through their postings or notes. \\

\subsubsection*{Network Creation: From interactions to edges}\label{Network_creation}

Both datasets contain pairwise interactions between users. Considering a dataset of $n$ users, we model each relation between user $i$ and user $j$ from such interactions as a random variable that follows a Bernoulli distribution with parameter $p_{ij}$.
We follow a Bayesian model using a beta prior for estimating $p_{ij}$ with parameters $\alpha_0,\beta_0$.
After observing all the interactions between $i$ and $j$ in the dataset, the posterior probability also follows a beta distribution, in this case parametrized by $\alpha_0+\text{pos},\beta_0+\text{neg}$, where $\text{pos}$ and $\text{neg}$ correspond to the number of positive and negative interactions respectively.

From these posterior probabilities, we build an undirected signed network $G=(V,E,\sigma)$, where $V$ is a set of $n$ nodes, $E$ is a set of $m$ edges, and $\sigma_{ij}$ is the edge sign. Edges are only defined for pairs of users who have a certain bias towards $0$ or $1$, i.e.,  $\mathbb{E}[p_{ij}]>0.6$ or $\mathbb{E}[p_{ij}]<0.4$, and very low uncertainty, i.e., $\text{Var}[p_{ij}]<10^{-4}$, where $\mathbb{E}[p_{ij}]=\frac{\alpha}{\alpha+\beta}$ and $\text{Var}(p_{ij})= \frac{\alpha \beta}{(\alpha+\beta+1)(\alpha+\beta)^{2}}$. For defined edges, we set their sign according to $\sigma_{ij}= \text{sign}\left(\mathbb{E}[p_{ij}] - \frac{1}{2}\right)$, i.e, two users have a positive (negative) edge if their expected posterior is above $0.6$ (below $0.4$) with high certainty.\\

\subsubsection*{Birdwatch}\label{Birdwatch}

Launched in January 2021, the platform aimed at fighting Twitter misinformation via crowd-sourced fact-checking by selected volunteer users, or \textit{birdwatchers}. These users assessed tweet trustworthiness with evaluative notes, including sources and arguments. The platform served as a small scale trial for the current known \textit{Community Notes}. Previous work analysis has shown high political alignment and polarization among users~\cite{prollochs2022community} and a tendency to scrutinize content from counter-partisans while following a partisan cheerleading behavior in ratings~\cite{allen2022birds}. 


Twitter regularly published updated and publicly available datasets containing metadata of notes (text, tweet ID, note timestamp, classification, note ratings) and anonymized users data. We retrieved all data covering the time span between the start of Birdwatch in January 2021 and August 2022.  Moreover, we re-hydrated the the content and metadata of the targeted tweets with the academic access to the Twitter API and computed an ideology score for the corresponding tweet author with Bayesian Ideal Point Estimation \cite{barbera2015tweeting} implemented by the package~\textit{tweetscores}. 


During the time span covered by our data, the platform changed their rating procedure from a simple \texttt{agree} versus \texttt{disagree} (Jan 2021 - Jun 2021) to \texttt{helpful}, \texttt{somewhat helpful} and \texttt{not helpful} for the remaining months in our data set (Jul 2021 - Dec 2022). Moreover, the platform launched a new algorithm to compute note statuses in February 2022, which searched for agreement across different viewpoints \cite{wojcik2022birdwatch}. Since these are substantial platform changes, we split the dataset into two parts accordingly: BW1, and BW2, and center our study mostly on BW1, leaving BW2 as comparison only since it includes a series of platform changes. BW1 includes $\sim 32k$ pairwise interactions between $2,676$ users, while BW2 is a larger dataset comprising $\sim 235k$ pairwise interactions amongst $10,662$ users.

On this platform, positive and negative interactions are present in similar proportions (see SI Appendix, Table S1). Both interactions in Birdwatch, agreement and disagreement, can be considered to be equally meaningful because they require an argumentation. Consequently, we use a uniform prior for the beta distribution that characterizes the user relations on Birdwatch. \\

\subsubsection*{DerStandard}\label{DerStandard}

The web page of the Austrian newspaper has a long tradition (dating back to the 1990's) of offering users discussion forums. 
Compared to other platforms with similar features, DerStandard uniquely provides information on which users rated a posting in addition to the sign of the rating (see SI Appendix, Fig. S1 for an example of the interface). A recent study shows that users that are active on DerStandard tend to be more often male, younger, more highly educated, and more often from Vienna or Upper Austria than respondents of a representative survey in Austria \cite{niederkrotenthaler2022mental}.

With permission from DerStandard, we automatically retrieved all publicly available postings and user ratings in the discussion forums below each news piece on DerStandard between Jan 2014 and Dec 2021. 
In addition to postings and ratings, we also retrieved tags that classify news pieces into topics according to the platform (e.g. sports, refugees in Austria, Op-Ed columns, etc).  

To analyze a stable user sample from DerStandard and avoid results originated from a large influx or outflux of users, we consider only users that rated at least once yearly in our observation period (begin of 2014 - end of 2021), thus removing accounts that have spurious activity levels. This allows us to identify roughly $14.8k$ users that we track over 8 years, comprising a total of $\sim 76M$ pairwise interactions. Our observation period spans major events, including the European refugee crisis (2015-16), the Austria government coalition dissolution due to corruption scandals (2019), and the COVID-19 pandemic (2020-21).


On DerStandard, negative interactions are underrepresented (SI Appendix, Table S1) and can carry a stronger signal than positive interactions. To account for that, we use a prior distribution that slightly favors negative interactions, especially when the volume of interactions is low, i.e., a beta distribution with $\alpha = 1$ and $\beta = 2$. The resulting network contains a similar number of negative and positive edges.

\subsection{Partitioning methods based on balance}\label{Network_theory} 
Our approach is based on finding the main division lines in a community and the posterior analysis of the reinforcement or challenge of those divisions. To find a robust partition based on balance, we build on previous work.

\subsubsection*{Main optimization problem}
A signed network is balanced if it can be partitioned into $k \leq 2$ groups such that all negative edges fall outside the groups defined by the partition and all positive edges fall within them. 
Following \cite{aref2018measuring} notation, given a signed graph $G=(V,E, \sigma)$, and a partition $P = \{ X, V \setminus X \}$, the frustration count will be the sum of the frustration state of all edges, $f_G(P) =  \sum_{(i,j)\in E} f_{ij}$, where $f_{ij}$ equals $1$ for frustrated edges and $0$ otherwise. Frustrated edges correspond to the edges that violate the assumptions of the optimal partition model, i.e. negative edges between members of the same group or positive edges between members of different groups. The problem thus is stated as finding the optimal partition $P^*$ with the minimum number of frustrated edges $ L_G^* = \min_P f_{G}(P)$. The value of $L^*_G$ can be used to compute partial balance.

\subsubsection*{Computational methods}\label{NPhard} The computation of $L^*_G$ is known to be NP-hard~\cite{aref2020modeling}. 
For small scale networks, however, exact computation of the frustration index is feasible using the binary linear programming formulation~\cite{aref2020modeling}. 
Several approximate methods have been proposed that are applicable to large scale networks. 
For example, Doreian and Mvar apply blockmodeling~\cite{doreian2009partitioning}, in which they optimize the criterion function $P(X) = E_{f,p} + E_{f,n} $ via a relocation algorithm, with $E_{f,p}$ defined as the frustrated positive edges and $E_{f,n}$ the frustrated negative edges. In practice, this method, combined with simulated annealing as in the \textit{Signnet} implementation~\cite{signnet}, provides approximate values of $L^*_G$ that correspond to robust partitions. Moreover, given that it involves a stochastic algorithm, we execute it $200$ times and select the partition yielding the minimum $L^*_G$ value. Further details regarding this approach can be found in the SI Appendix, section 2A.
Any approximated value for $L^*_G$ will necessarily be equal or higher than its exact value, given that there is no sub-optimal partition that can provide a smaller number of frustrated edges, thus the best approximated value will be an upper bound.

\subsubsection*{Generalization to more than two groups}
All the previous definitions and methods are generalizable to $k>2$ partitions \cite{davis1967clustering,aref2021identifying}, which corresponds to a definition of weak structural balance. In that case, each value of $k$ provides an optimal solution $L^*_{G} (k)$, and a reasonable selection is to keep the value of $k$ which yields the minimum $L^*_G$. In \cite{doreian2009partitioning}, it is shown that $L^*_G$ follows a concave curve with a unique minimum value of $k$, which we refer to as $k^*$. For all the case studies presented in this paper we find $k^* = 2$, ergo two differentiated factions. See SI Appendix, section 2B for details in the multi-partition selection for our data.

\subsubsection*{Previously defined useful metrics}

Given a partition, a metric that can be useful in determining the incidence of the general division on the community, or the degree up to which the network can be easily separated into groups, is the "Normalized Line Index of Balance" \cite{aref2020multilevel}: $1 - \frac{L^*_G}{m/2}$. The normalizing factor $m/2$, where $m$ is the volume of edges in the network, accounts for different network sizes and is a soft upper bound on the number of frustrated edges. This is a partial balance index that ranges from low to high balance, with 1 being the completely balanced case. Note that the index decreases the more frustration there is (i.e. the more blended the groups are). Therefore, when working with an approximated $L^*_{G}$, the index value is a lower bound. 

Interestingly, we can understand this level of grouping as a structural measure related to polarization under the assumption that there is a small number of groups which are of similar sizes. Otherwise, it would either be a a complete fragmentation into small groups or a case of a majority versus minorities. We verify this assumption in the \hyperref[Results]{Results} section. 

Similar to earlier work~\cite{aref2020multilevel}, we can also analyse the two mechanisms that are involved in the coherence of users's links to the partition: alignment with one's own group (Cohesiveness) and alignment against the opposing group (Divisiveness). 
Cohesiveness (Divisiveness) is defined by the proportion of internal (external) edges that are positive (negative). Given our optimal partition $P^*$, internal edges are defined by $E_{p}^{i} = \{ (i,j) \in E | i,j \in X$ or $i,j \notin X \}$ and external edges are defined by $E_{p}^{e} = \{ (i,j) \in E | i \in X, j \notin X $ or $ j \in X, i \notin X\}$. 

\begin{figure*}[h!]
\centering
\includegraphics[width=0.48\linewidth]{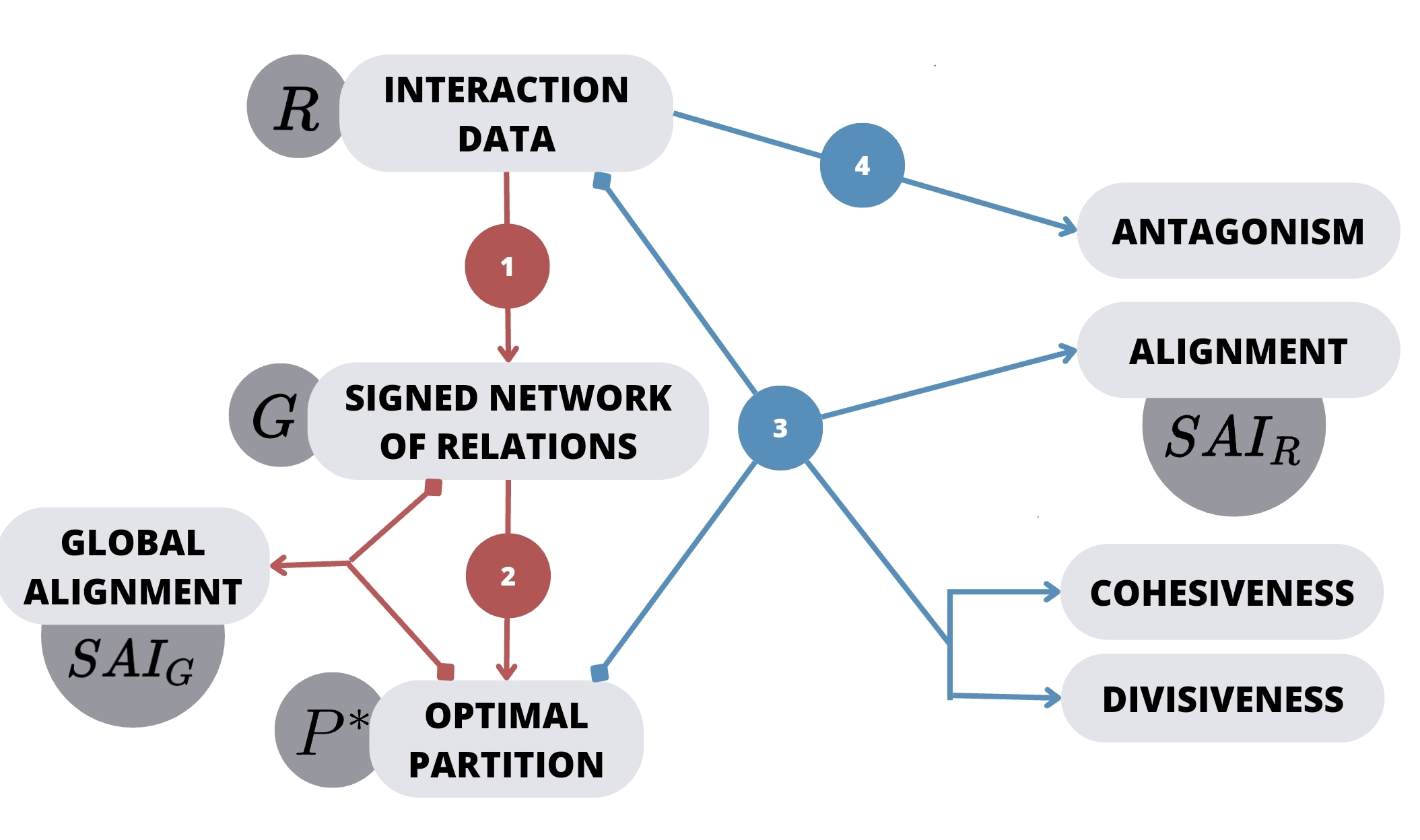}
\includegraphics[width=0.45\linewidth]{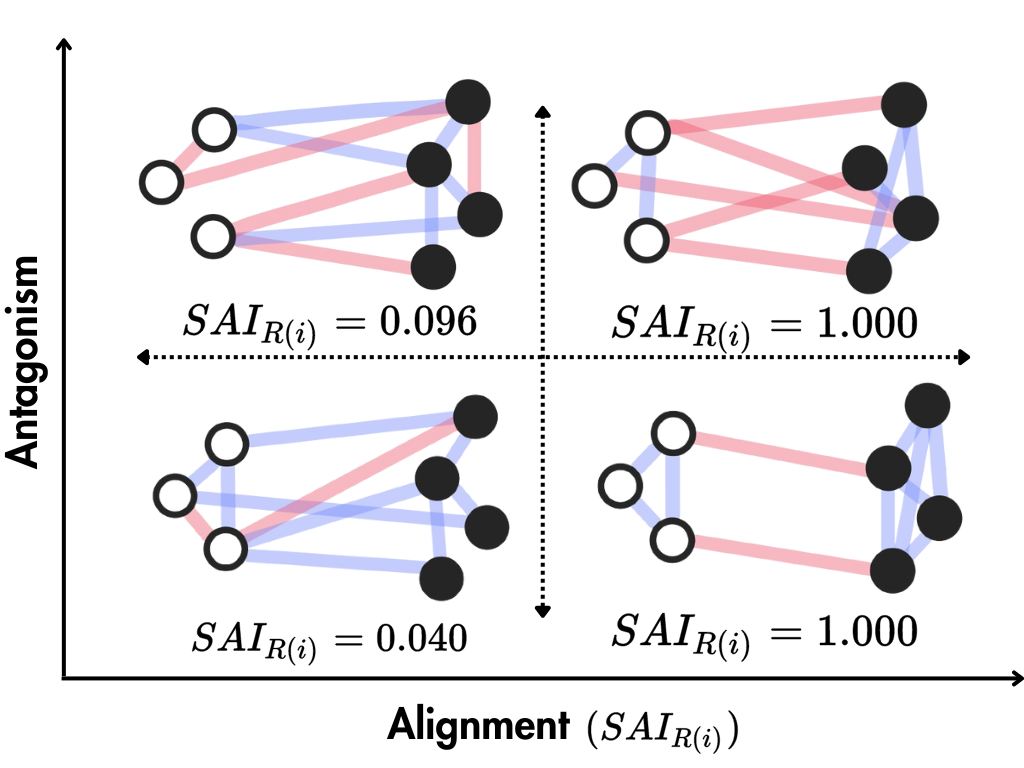}
\caption{\textbf{(Left) Schema of the FAULTANA pipeline: our analysis framework for Antagonism, Alignment, Cohesiveness, and Divisiveness.} Grey boxes indicate data structures and variables implicated in the pipeline. Step 1 creates the relation network based on an aggregation of interactions through time. Step 2 applies the optimization algorithm, either exact or approximated, to obtain the optimal number of groups and optimal partition. From these two steps we can retrieve a global Alignment metric $SAI_G$. Then, by selecting subsets of the interaction data and with the optimal partition information, steps 3 and 4 compute the four metrics of interest: Antagonism, Alignment, Cohesiveness (normalized) and Divisiveness (normalized). \\
\textbf{(Right) Illustration of Antagonism and Alignment in signed network examples.} The four networks have been constructed with the same edge density, number of nodes, and partitioning of nodes. Negative edges are red and positive edges are blue. The two upper networks have a higher proportion of negative edges, and thus higher Antagonism than the ones on the lower quadrants. Computed $SAI_{R(i)}$ values are provided to illustrate that the two right quadrants exhibit a higher level of Alignment, which is due to the lower amount of frustrated edges. Only the right upper quadrant corresponds to a strict definition of polarization in terms of both Antagonism and Alignment. }
\label{fig:pipeline}
\end{figure*}

\subsection{The FAULTANA Pipeline}\label{faultana}

After constructing the signed relation network of each platform, we obtain their optimal partitions using the methods described above. Once the belonging of users to each group is fixed, we can assess the status of the platform globally (network of relations) or describe the status of directed sub-sets of the data (network of interactions). These allow us to find four metrics of interest: \textit{Alignment}, \textit{Antagonism}, \textit{Cohesiveness} and \textit{Divisiveness}. We designate this set of steps as the FAULTANA pipeline, which stands for FAULT-line Alignment Network Analysis (See Figure~\ref{fig:pipeline}).

\subsubsection*{Re-normalization and global metrics}\label{renorm} 

In order to be able to compare across subsets of our data, we have to re-think some of the characteristics of the previously designated metrics like the "Normalized Line Index of Balance", Cohesiveness or Divisiveness.

In the case of the "Normalized Line Index of Balance", we re-normalize it by comparing the empirical estimate of $L^*$ versus its mean value in repeated measurements of a null model. The null model based on graph $G$ randomly re-distributes sign attributes while keeping the overall structure of the network and the partition fixed ($\widetilde{G}$). 
The value of $L$ in the null model simulations, $L_{\widetilde{G}}$, is consistently higher than the frustrated edges in our datasets, proving to be a tighter bound than only considering the number of edges with the term $m/2$. 
Thus, we define the \textit{Global Signed Alignment Index}, which we also simply call \textit{Global Alignment}, as: 

\begin{figure}[h!]
\begin{align}
SAI_G&=\langle 1 - \frac{L^*_G}{ L_{\widetilde{G}} } \rangle 
\end{align}
\label{eqn:SAIG} 
\end{figure}

We obtain $95\%$ confidence intervals for $SAI_G$  from the distribution resulting of repeated instances of the null model.

On the other hand, the measures of Cohesiveness and Divisiveness defined above cannot be compared between systems with different ratios of negative versus positive interactions. For example, a system A with a higher ratio of negative interactions than system B will have by construction a higher Divisiveness even if it is not more strongly divided along the fault line than system B. This can be observed in simulations of our null model, which show that the expected value of Divisiveness and Cohesiveness is perfectly correlated with the fraction of negative interactions (Fig.S4). To solve this, we design new metrics of Normalized Divisiveness and Normalized Cohesiveness by subtracting from the original metrics (i.e. not normalized) the mean values obtained in the null model simulations. For brevity, we will use the original metric names to refer to their normalized versions throughout the remainder of the manuscript. We assess the uncertainty of our measurements of Divisiveness and Cohesiveness through bootstrapping. For each measurement, we create $10,000$ bootstrap samples of the network with replacement and of the same size as the original. On each bootstrap sample, we calculate Divisiveness and Cohesiveness and we use the resulting values to calculate the bootstrapping confidence interval around our original measurement.

\subsubsection*{Formalization of Alignment} Our normalization approach allows us to obtain a meaningful $SAI$ for sub-sets of the interaction data. To do so, we maintain the optimal partition obtained from the network of relations (i.e. we fix the belonging of each user to a group that is defined by the long-term relation between users), and we proceed to assess how aligned the interactions within that subset of the data are to this partition. We refer to this metric as \textit{Alignment}, or $SAI_{R}$. Since it follows the same laws of frustration (e.g. negative interactions within a group are frustrated interactions, and so on), we just have to re-define the $SAI_G$ in the following way: 

\begin{figure}[h!]
\begin{align}
SAI_{R}&=\langle 1 - \frac{L_{R}}{ L_{\widetilde{R}} }\rangle 
\end{align}
\label{eqn:SAIR} 
\end{figure}

where $R$ accounts for the network of directed interactions within a set or subset of the data, denoted by $R(t)$ in case of a temporal subset, or $(i)$ for a selection based on issue or topic. $L_{R}$ is then the number of frustrated interactions in that network given the existing assignment of nodes to groups. As in the case of $SAI_G$, $\widetilde{R}$ denotes an instance of the null model applied on $R$, by reshuffling the sign configuration while keeping the network structures and groups. 

\subsubsection*{Formalization of Antagonism} Additionally, we formally describe \textit{Antagonism} as the proportion of negative interactions in $R$, which is a simple indication of the prevalence of conflict or general disagreement. This measure is then not related to the network structure, like Alignment, but it indicates a property of the user-content interaction in terms of the overall presence of disagreement in comparison to agreement.

\subsubsection*{Formalization of Cohesiveness and Divisiveness (local)} Besides computing Cohesiveness and Divisiveness for the network of relations, providing a general overview of each platform, we can also analyse these metrics for subsets of interactions associated with topics or time periods, which can show how Cohesiveness and Divisiveness contribute in Alignment changes. Furthermore, given the directionality of ratings in the network of interactions, we can examine group asymmetries by calculating the separate contribution from each group to these metrics (e.g. how much of the division between two groups is driven by one of them).

\subsubsection*{Conceptualization of Alignment and Antagonism}\label{alignment} 
The metric of \textit{Alignment} captures how interactions follow the division of the network into opposed groups, while our metric of \textit{Antagonism} captures the overall tendency towards negative interaction in the network regardless of groups.
By considering both these measures, we can provide a more comprehensive picture of polarization than when these two concepts are not explicitly distinguished. Figure~\ref{fig:pipeline} shows how these two metrics capture various polarization scenarios given a partition of the network into groups and the positive and negative interactions in the system. A network with low Alignment and low Antagonism has few negative interactions and no strong division into groups, corresponding to a situation with the weakest polarization. The lower right part of the space, where Alignment is high but Antagonism is low, corresponds to an \textit{echo chamber} case in which most interactions are positive but happen between like-minded individuals and not across groups.  The upper left cases are networks with high Antagonism but low Alignment, capturing scenarios where disagreement exists but not necessarily following the division of the network into groups. This can happen when everyone is against everyone or where other divisions exist but do not follow the general ideological separation of the network into groups. And finally, the upper right part of the space corresponds to cases where polarization is high, as both Antagonism and Alignment are high. In this high-polarization case, there is a strong cleavage between groups such that positive interactions are confined within groups while frequent negative interactions happen mostly across groups.

\section{Results}\label{Results}

\subsection*{Approximating Alignment in Birdwatch}\label{Birdwatch convergence}

In this section, we evaluate our methodology and its performance based on the results obtained from the two Birdwatch datasets. We use Birdwatch for two key factors. Firstly, as described in Section \hyperref[NPhard]{Computational methods}, we can run the exact method for small networks, while for large networks we have to run the approximate algorithm due to the complexity of the problem. The size of BW1 allows us to run both the exact and approximate algorithms and compare the solutions to estimate the difference in signed networks of this kind. The results of both algorithms are very similar in BW1, with the approximated $SAI_G$ being $84\%$ of the $SAI_G$ obtained with the exact method and an average partition overlap coefficient~\cite{vijaymeena2016survey} of $0.89$.

\begin{figure}[t!]
\centering
\includegraphics[width=0.95\linewidth]{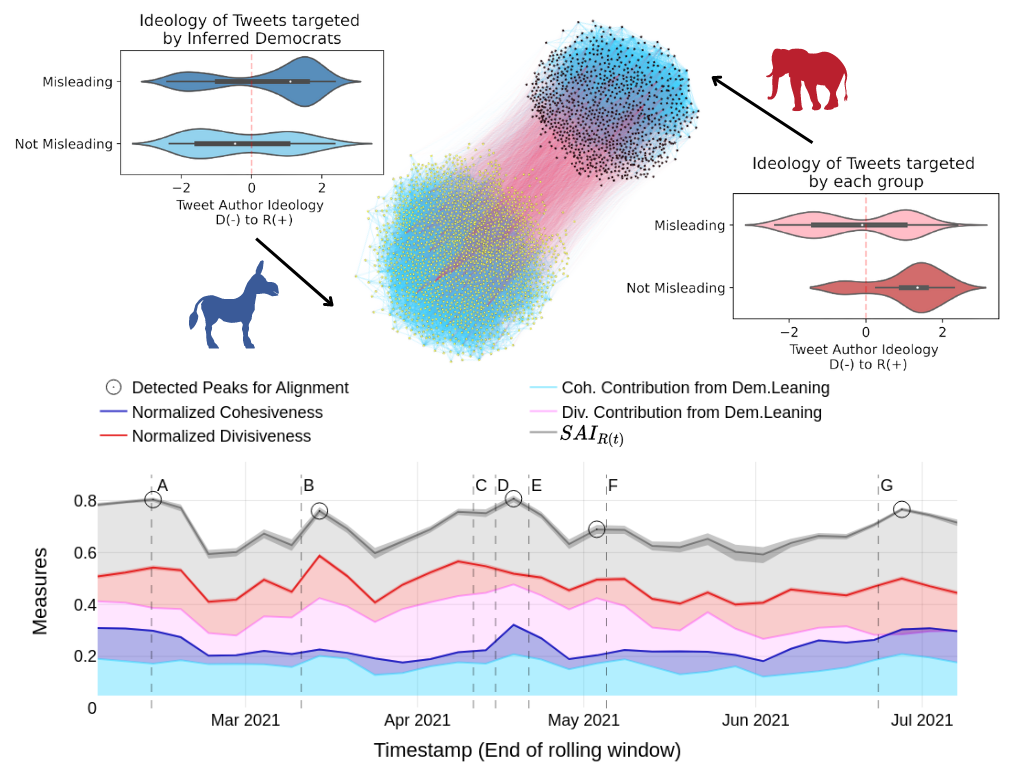}
\caption{\textbf{(Upper figure) Signed network visualization of Birdwatch.} Network of signed relationships for the BW1 dataset, comprising a total of 2,676 users and around 25,562 edges, negative colored red and positive blue. Node color corresponds to their group membership as identified by the exact method. Nodes belonging to the largest (smallest) group are depicted in yellow (black). Negative edges tend to connect different groups, while positive edges predominantly connect nodes within groups, demonstrating a considerable degree of balance. \\
\textbf{Insets:} Inferred ideology of the targeted tweet's author separated by which group targeted the tweet and the nature of the note. We can only retrieve a score for tweet authors that have connections to political actors ($\sim 60\%$ of the users that posted tweets targeted in Birdwatch).
The larger group gives misleading notes with more probability to tweets authored by Republican users, i.e. counter-partisan policing, with a slightly higher tendency to give not misleading notes to tweets by Democrat users. Thus, we identify the larger group as Democrat-leaning. The smaller group is much more likely to give not misleading notes to tweets authored by Republican users, showing a pattern of cheer-leading within Republicans and thus being identified as Republican-leaning. \\
\textbf{(Lower figure) Timeline of Alignment, Cohesiveness and Divisiveness in Birdwatch (BW1).}
The time series of each metric is calculated over a rolling window of ten days with increases of 5 days, with values allocated on the right of each window.  The shaded area around Alignment time series shows 95\% Confidence Intervals calculated against $10,000$ instances of the null model. 
Divisiveness is shown in red and Cohesiveness is shown in blue, with lighter areas showing the contribution of Democrat-leaning users to each metric and the remaining area above showing the contribution of Republican-leaning users.
Bootstrapping intervals in Divisiveness and Cohesiveness are obtained for 10,000 bootstrap samples with replacement. 
The Alignment measure, $SAI_{R(t)}$, oscillates around a mean value of $0.65$. Divisiveness stays consistently above Cohesiveness, showing that negative interactions are the main driver of Alignment. Detected peaks in $SAI_{R(t)}$ are marked with circles and notable political events in the US are marked with vertical dashed lines for reference. For each peak, a summary text analysis of tweets in that period is shown in  SI Appendix, Table S4, which can be further contextualized as increases in Cohesiveness, Divisiveness, or both. 
An interactive version of this plot can be found at \href{https://emmafrax.github.io/BW1.html}{https://emmafrax.github.io/BW1.html}
}\label{fig:bw_panel}
\end{figure}

For both BW1 and BW2, we find that the optimal number of groups is $k^* = 2$, and the largest groups contain roughly twice the number of users of their smaller counterparts (see SI Appendix, Table S2 for more details). Figure~\ref{fig:bw_panel} shows the signed network of relations obtained from BW1. Previous literature focused on Birdwatch suggests that the platform is characterized by two opposing factions, corresponding to Republican- and Democrat-leaning users, who attach notes to tweets following behaviors of \textit{counter-partisan policing} and \textit{inner-partisan cheer-leading} \cite{allen2022birds}. By building on the ideology score extracted from the tweets, we test whether the groups identified through our method reproduce this behavior, thereby evaluating the coherence of our approach with other metrics of political alignment.

When we retrieve the notes that users from each of these partitions have given to tweets, we find evidence of these policing-cheerleading patterns, as our largest group - which we denote as \textit{inferred Democrats} - is strongly biased towards tagging Republican-leaning tweets as misleading. Contrarily, the smaller partition - \textit{inferred Republicans} - consistently rates like-minded tweets as not misleading (see Figure~\ref{fig:bw_panel}).

\subsection*{Evolution of Alignment in Birdwatch}\label{Birdwatch Results}

The lower part of Figure~\ref{fig:bw_panel} shows the time series of $SAI_{R(t)}$ in BW1. The fluctuations in the measure over time indicate whether the level of Alignment among interactions increased or decreased during that particular period. The time series of normalized Cohesiveness and Divisiveness contextualize these movements, as they show whether peaks are due to higher cohesion within groups or higher division between groups, and what is the contribution of each group to these metrics. In these time series, Antagonism and Alignment have a low correlation, which emphasizes the need to consider them as two different measures (more details can be found in SI Appendix, Fig. S5).

We applied a peak detection algorithm and identified five local maxima of $SAI_{R(t)}$ that are marked in Figure~\ref{fig:bw_panel}. To understand the context of the tweets on the day of each peak, we generated wordshift diagrams~\cite{gallagher2021generalized} for each peak in comparison to the rest of the tweets. Details on the wordshift diagrams can be found in SI Appendix, Fig. S7, Fig. S8 and Fig. S9. 
Our analysis shows that peaks of Alignment happen around controversial topics in the US. For example, we see that the second peak, associated with events related to  COVID-19 vaccination (B),  is driven by an increase in Divisiveness, especially from Democrat-leaning users. Alternatively, the third detected peak, which is associated with events about police shootings (C,D,E) , has a stronger contribution of Cohesiveness, especially within the Republican-leaning users. The other three peaks (1st, 4th and 5th) are driven by a mix of Cohesiveness and Divisiveness. The keywords and events at those time periods point towards discussions regarding the US Government and its policies (G), Donald Trump and 2020 election results (F), and other relevant events such as the Capitol insurrection or the Texas Power Crisis (A). A list of keywords and identified relevant events can be found in SI Appendix, Table S4.

\subsection*{Results for DerStandard}\label{DerStandard Results}

\begin{figure}[t!]
\centering
\includegraphics[width=0.48\linewidth]{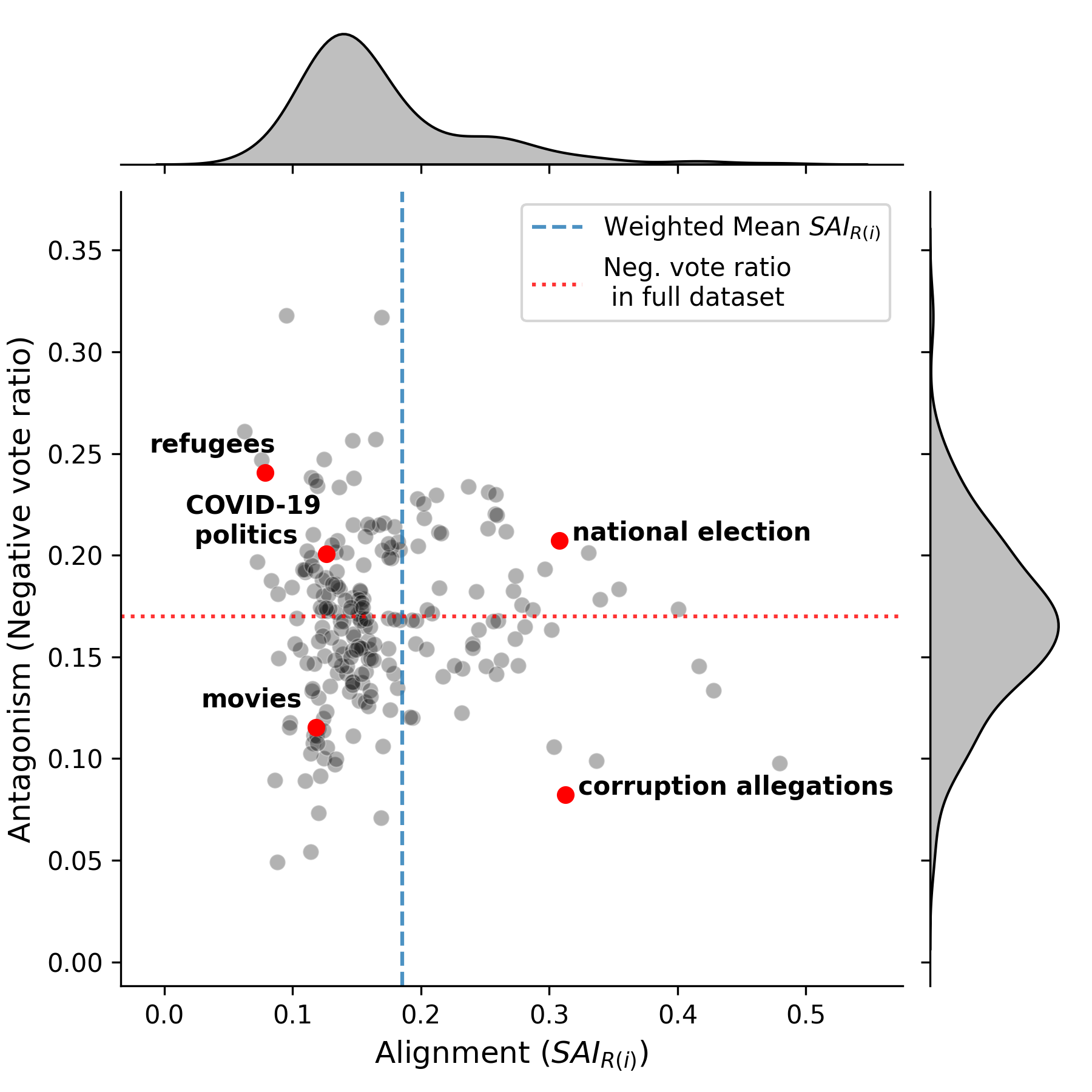}
\includegraphics[width=0.48\linewidth]{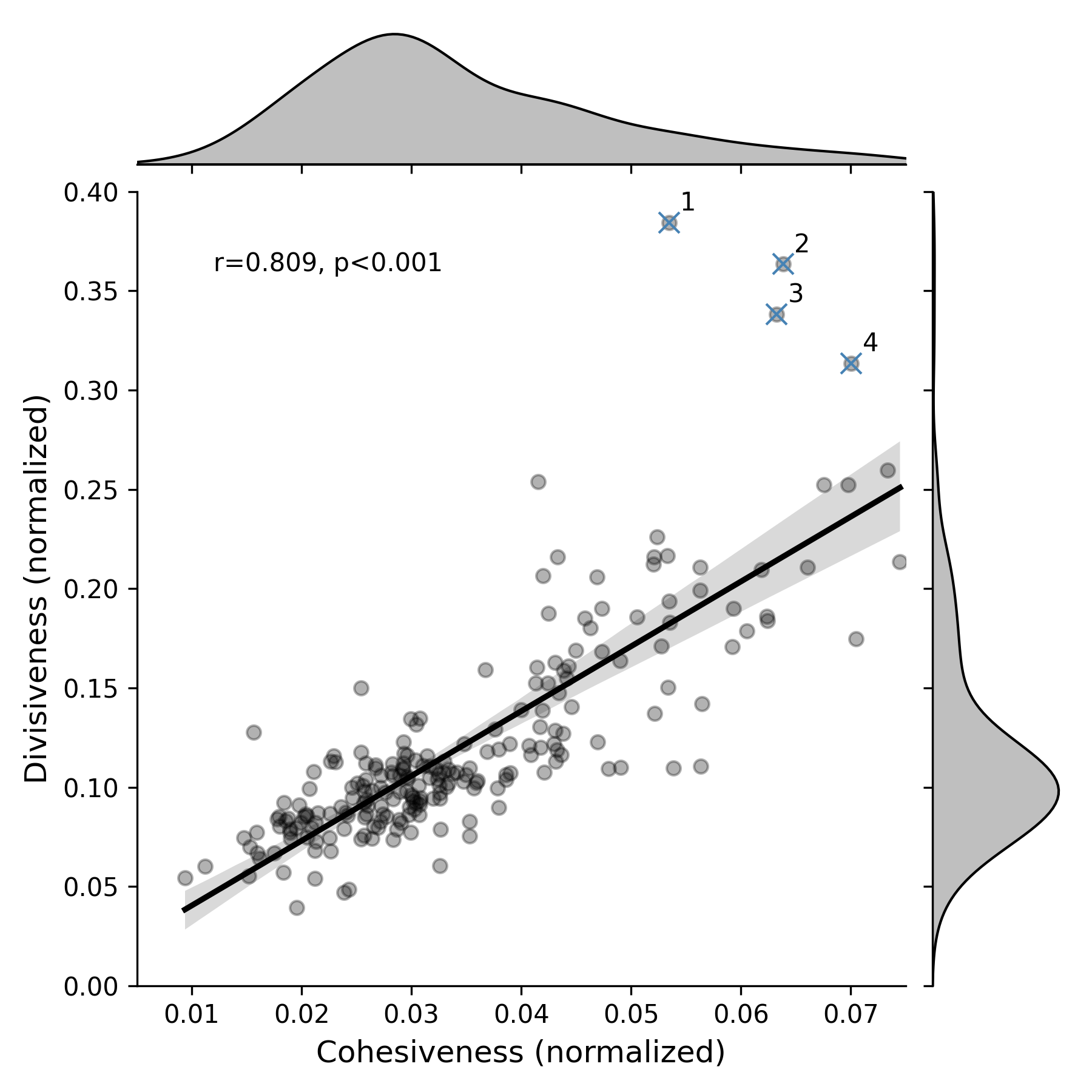}
\caption{\textbf{Alignment versus Antagonism and Cohesiveness versus Divisiveness across DerStandard topics.}
The left figure shows Antagonism and Alignment of the ratings of each news topic in DerStandard. Topics have been selected based on the topic/subtopic tags associated with the articles located above the postings (e.g., sports, climate change, etc.). Dashed lines show the mean values of each metric to identify the quadrants depicted in Figure~\ref{fig:pipeline} 
An interactive version of this figure can be found at  \href{https://emmafrax.github.io/scatter.html}{https://emmafrax.github.io/scatter.html}.
The right figure shows the scatterplot of Divisiveness versus Cohesiveness for DerStandard rating sub-sets based on topics. These two measures, which account for two different mechanisms that define Alignment, have a significant correlation across topics of 0.8. The highlighted outliers correspond to: (1) BVT (Austrian counterterrorism agency), (2) Abortion, (3) Scheuba (Austrian comedian) and (4) ÖVP (Political Party)}\label{fig:der_scatter}
\end{figure}

Our approach to detecting groups in the DerStandard network shows that this network has an optimal $k^*$ of two groups, as in Birdwatch. The size of these groups is slightly more similar, with the largest one comprising $62\%$ of the nodes. Even though the DerStandard dataset spans a much longer period and contains more users than the Birdwatch datasets, the Global Alignment of the network is substantially high ($SAI_G=0.3955$), showing that alignment can appear across different sizes and time scales. Divisiveness ($0.2899$) is still considerably higher than Cohesiveness ($0.1409$), also mirroring the results for Birdwatch. More details on these results can be found in SI Appendix, Table S3.

Given the classification of news in DerStandard, we can measure Alignment and Antagonism on the set of user ratings focused on specific topics, thus locating issues in the space of network structures shown in Fig. \ref{fig:pipeline}. The scatter plot for Alignment and Antagonism of DerStandard topics is shown in Figure \ref{fig:der_scatter}, where the spread of values allows for all four combinations outlined by our approach. Alignment and Antagonism have a low correlation across topics ($r = -0.0016, p = 0.981, 95\%$ CI $[-0.134,0.131] $), suggesting that these two concepts should not be conflated into a general dimension of polarization. By inspecting the topics falling into each quadrant of the plot, we find their distribution agrees with intuitive expectations. For example, topics with a high conflict potential such as migration, COVID-19 politics, gender politics, climate change, and elections are on the high range of Antagonism, whereas lifestyle, sports, and culture topics such as movies, family, travel, art market or international football are located in the low ranges of Antagonism. With regard to the dimension of Alignment, we find that conflicting topics such as national elections, abortion, military service, or climate change are more aligned than migration or COVID-19 politics. These last two were indeed issues that did not divide the Austrian population clearly into left and right.  Note that theselower patterns cannot be explained by the number of ratings, posts, or articles on each topic, as shown more in detail in the SI Appendix, section 3B. 

We highlight a few examples within each quadrant of Figure \ref{fig:der_scatter} to better illustrate how Alignment and Antagonism relate to each other. While \textit{Refugees} and \textit{COVID-19 politics} are identified as conflicting topics, resulting in higher levels of Antagonism, they do not align precisely with the primary division line. During the crucial years for those topics of 2015/16 and 2020/21, we have seen some unexpected political alliances that do not follow from a classical left-right spectrum. These include common platforms between the anti-migration left and right-wing populists or the anti-statist right and anti-vaccine parts of rather left-wing Green parties. These agreements on certain issues between otherwise ideologically distant parties have historically been described by the term "Querfront" ("cross-front") \cite{dewiki:236477099}. Conversely, the tag \textit{National elections} exhibits both Antagonism and Alignment, indicating a combination that favors polarization. This can be explained by federal elections in a representative democracy to lead to more discussion along traditional party lines. Additionally, \textit{Corruption allegations} pertains to specific events involving some of the political parties in Austria. Although it demonstrates Alignment, these particular events did not generate substantial conflict within the platform. This could potentially be due to a limited number of defenders of those specific parties that have been covered much in the news in a corruption context (FPÖ and ÖVP, resulting from their joint government coalition), as DerStandard is historically considered a more left-liberal-leaning newspaper. As expected, a more offtopic tag such as \textit{Movies} exhibits low levels of both Alignment and Antagonism.

While Antagonism and Alignment across topics are weakly correlated, Cohesiveness and Divisiveness are strongly correlated, as shown on the lower panel of Fig \ref{fig:der_scatter}. This is expected, as the affective component of polarization captured by Alignment implies a correlation between out-group animosity and in-group support. Nevertheless, there are topics that deviate from the association between Cohesiveness and Divisiveness by having substantially higher Divisiveness: BVT (Institution), Abortion, Scheuba (Austrian comedian) and ÖVP (Political Party) (see lower panel of Fig \ref{fig:der_scatter}), while this pattern is not mirrored for Cohesiveness. As with the time series of Alignment on Birdwatch, measuring Cohesiveness and Divisiveness is informative even though they both form part of the same phenomenon of Alignment.

\begin{figure}[]
\centering
\includegraphics[width=0.98\linewidth]{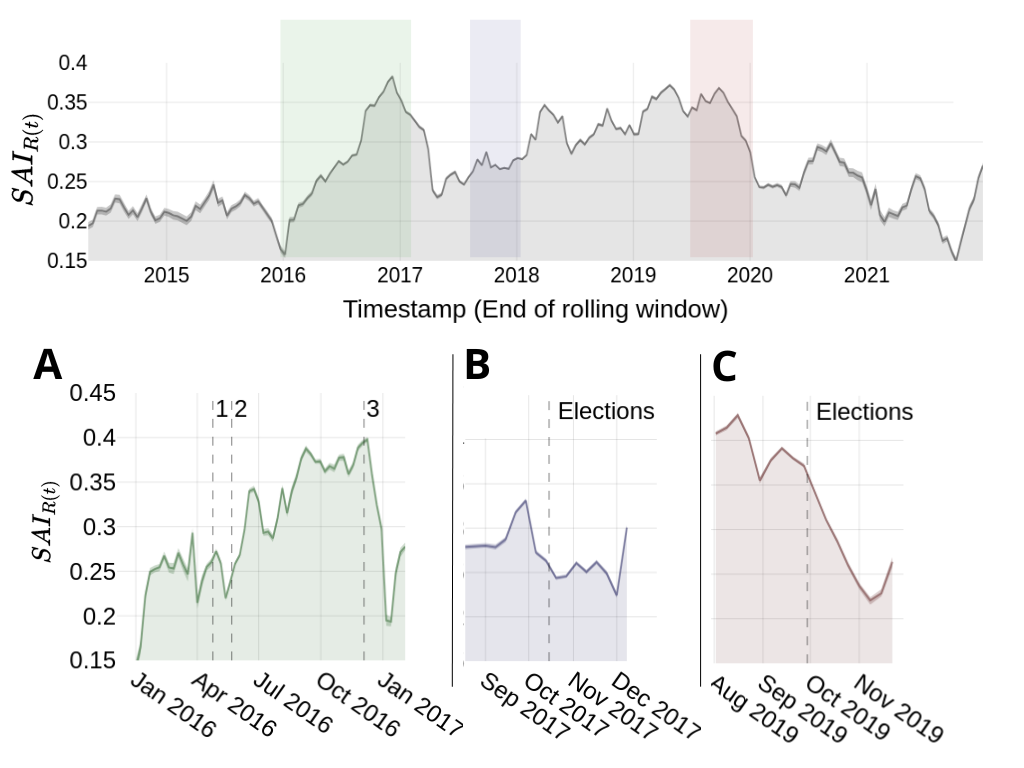}
\caption{\textbf{Alignment timeline in DerStandard ratings sub-set of political topics, with detailed fluctuations in election periods.} Upper timeline figure shows the Alignment measure obtained using a rolling window of 120 days of width and a step of 14 days. The features of the rolling window are selected so that the trends in Alignment through the eight years are visible, e.g. the change in trend at the start of 2016. In the lower figures we show more detailed changes of Alignment, with a rolling window of 30 days of width and a step of 7 days, around the three repetitions of the 2016 Presidential elections (A: 1,2 and 3) and the 2017 and 2019 Legislative Elections (B and C). }
\label{fig:der_timeline}
\end{figure}

The time series of Alignment in DerStandard reveals how cleavages become salient around politically-relevant events. 
Figure~\ref{fig:der_timeline} shows the time series $SAI_{R(t)}$ for all DerStandard discussions in news on three topics: national elections, parties, and the federal president. This highlights political discussions from other, less-contentious topics as identified above.
There is a clear change in the trend of Alignment at the beginning of 2016, showing steady growth up to the beginning of 2017.
This falls into the time period of the so-called "2015 European migrant crisis" \cite{2015EuropeanMigrant2023} when migrants arrived in Europe in numbers that were unprecedented since World War Second. While migration started before 2016, the rise in Alignment starts right after the reporting of sexual assaults during New Year's Eve 2015-2016 celebrations in Cologne, Germany\cite{201516New2023}, which were widely covered in German-speaking media and debated over the following year.

Political events can also drive decreases in Alignment, especially if we consider that Austria has a multi-party system. After an election, the political climate changes toward building government coalitions with multiple parties, thus predicting lower Alignment as suggested by the case of online networks of Swiss politicians \cite{garcia2015ideological}. This can be observed in the time series of Alignment in DerStandard if we zoom in to recent elections. 
Panel A of Figure \ref{fig:der_timeline} shows the timeline of Alignment during 2016, where the increase in Alignment that year accelerates after the result of a presidential election was overturned by the Supreme Court of Justice. This controversial decision lead to a period of increased Alignment towards the repetition of the election, to then quickly reset to earlier levels of Alignment as soon as the repeated election took place and a candidate won by a large margin.

Panel B of Figure \ref{fig:der_timeline} shows a decrease in Alignment that happened shortly before the 2017 legislative elections, which was called early since there were clear favorite parties to form a coalition in pre-election polls.
The effect of the legislative elections in 2019 (panel C) showed a sharp decrease in Alignment afterward, as the result was not as clearly expected as in 2017 and which led to a new government coalition with a party that was not involved in the previous government.


\section{Discussion}\label{Discussion}

We have successfully factored online polarization into two dimensions: Antagonism, representing the level of hostility or disagreement in an online discussion, and Alignment, determined by the tendency of individuals to position themselves in a discussion according to their belonging to a group (or for that matter, their positioning across the "other" group(s)). These two measures, although both contributing to polarization, have distinct characteristics and are weakly correlated across topics. We discovered that large-scale online political discussions exhibit an underlying polarized structure based on balance, which becomes more prominent when examining discussions centered around aligned topics. An essential takeaway is that online polarization is a dynamic, responsive phenomenon deeply influenced by contemporary political and societal events. It exhibits rapid responses, but through an examination spanning a sufficiently extended time frames, such as in the case of DerStandard, we can discern overarching trends alongside specific peaks. 

Particularly, in terms of insights drawn from the study of Birdwatch, we found that changes in polarization can arise from different mechanisms (i.e. Cohesiveness or Divisiveness) within one or both of the groups. Additionally, the identification of Republicans and Democrats provides valuable insights into the status of each topic and positioning in relevant online discussions. Our findings on Birdwatch, as a platform dedicated to crowd-sourced fact-checking which has now been extended globally, can be beneficial to understanding the dynamics and effectiveness of using a wisdom-of-the-crowds approach to combat misinformation.

On the other hand, through our comprehensive analysis of DerStandard, we have uncovered that topics such as COVID-19 politics and Refugees, despite their contentious and relevant nature in online discussions, do not align with Austria's general Left-Right divide. This finding sheds light on the political divisions within Austria and serves as evidence that our methodology is capable of identifying cross-cutting cleavages, as these are topics with high antagonism but lower alignment. Furthermore, through an analysis of the temporal trends of Alignment pertaining to politically relevant topics, our findings demonstrate coherence with expected behaviors given the context of the respective time frames.

FAULTANA, our proposed pipeline, is agnostic in terms of political system (is applicable to multi-party as well as two-party), language, or issue dimensions, and can be extended to other use cases as long as positive and negative interaction information is available. It can also be tuned to platform-specific features, for example choosing the prior distribution for user relations. In the specific cases of DerStandard and Birdwatch, we were able to retrieve a division in the ideological spectrum (left .vs. right), but it is possible that other platforms' main divisions fall between other social, demographic or ideological positionings. Therefore, it allows us to study the main cleavage in a platform's community without the need of classifying users by their opinions \textit{a priori}. 

Our work is subject to several limitations, the main one being that we have to use approximated methods to find (near)-optimal partitions for large scale networks. However, even in that situation, we still capture significant values for our metrics and our approximated results are comparable to the exact results for the BW1 dataset, which brings us to the conclusion that we are still measuring what we aimed to, even if not at the highest accuracy possible. Moreover, even in the exact solution it is not possible to ensure a unique single optimal partition, since the method only ensures a unique solution for the minimum amount of frustrated edges, and several partitions can satisfy that requirement~\cite{aref2018measuring}.

On the other hand, a relevant assumption is on the fixed belonging of users to a group defined by the optimal partition. We are assuming there is a global clustering to which users are aligned. This is not too far-fetched given the fact that there tends to exist issue alignment in society \cite{baldassarri2008partisans,dellaposta2015liberals} and we consider different numbers of clusters that lead to lower alignment. However, for long time scales or unprecedented global phenomena such as the COVID-19 pandemic, there could be relevant changes in these structures. 

We leave for future work the inclusion of a tracing system that assesses the partition quality through time and updates it accordingly. With access to longitudinal data for large populations, our method could be very useful to automatically detect shifts in the main lines of division, which would provide more accurate pictures of polarization in terms of its components of alignment and antagonism, how it manifests across topics, and how it evolves over time.

\backmatter

\bmhead{Supplementary information}

This article has an accompanying supplementary file.

\bmhead{Acknowledgments}

We thank Samin Aref for insightful discussions on the methodology.

\section*{Declarations}

\begin{itemize}
\item D.G. and M.P. acknowledge funding from the Vienna Science and Technology Fund (WWTF) [10.47379/VRG16005]
\item The authors declare no conflicts of interest.
\item Ethics approval: all analyses were observational and based on publicly available data archives in Birdwatch and DerStandard. The study contained no interaction with any of the users of either platform.
\item \textbf{Data availability.} In order to ensure the reproducibility of these analysis and contribute to the field with our curated dataset, we intend to release the network data for Derstandard once the peer-reviewed version of the article is published. Until that time, we are willing to share the data upon request. The individual responsible for inquiries regarding the data is the corresponding author, EF.
\item \textbf{Code availability.} In order to ensure the reproducibility of these analyses, we intend to make the code publicly available once the article is published. Until that time, we are willing to share the code upon request. The individual responsible for inquiries regarding the code is the corresponding author, EF.
\item \textbf{Authors contributions.} E.F. carried out the analyses, code generation, visualizations and main draft for the manuscript. M.P retrieved the data, helped with the analysis and provided contextualization for DerStandard. S.S. provided his insights and expertise in the study of polarization and political systems, and embedded our findings in the relevant literature. V.G. supervised and assessed the methodology approaches and provided original ideas for the formal definition of our metrics. D.G. provided global guiding and supervision through the project and several of the ideas behind this framework. All authors contributed to writing the manuscript, and took part in the discussions and decisions which resulted in this work. 
\end{itemize}

\noindent


\bibliography{sn-bibliography} 


\begin{thebibliography}{54}
\ifx \bisbn   \undefined \def \bisbn  #1{ISBN #1}\fi
\ifx \binits  \undefined \def \binits#1{#1}\fi
\ifx \bauthor  \undefined \def \bauthor#1{#1}\fi
\ifx \batitle  \undefined \def \batitle#1{#1}\fi
\ifx \bjtitle  \undefined \def \bjtitle#1{#1}\fi
\ifx \bvolume  \undefined \def \bvolume#1{\textbf{#1}}\fi
\ifx \byear  \undefined \def \byear#1{#1}\fi
\ifx \bissue  \undefined \def \bissue#1{#1}\fi
\ifx \bfpage  \undefined \def \bfpage#1{#1}\fi
\ifx \blpage  \undefined \def \blpage #1{#1}\fi
\ifx \burl  \undefined \def \burl#1{\textsf{#1}}\fi
\ifx \doiurl  \undefined \def \doiurl#1{\url{https://doi.org/#1}}\fi
\ifx \betal  \undefined \def \betal{\textit{et al.}}\fi
\ifx \binstitute  \undefined \def \binstitute#1{#1}\fi
\ifx \binstitutionaled  \undefined \def \binstitutionaled#1{#1}\fi
\ifx \bctitle  \undefined \def \bctitle#1{#1}\fi
\ifx \beditor  \undefined \def \beditor#1{#1}\fi
\ifx \bpublisher  \undefined \def \bpublisher#1{#1}\fi
\ifx \bbtitle  \undefined \def \bbtitle#1{#1}\fi
\ifx \bedition  \undefined \def \bedition#1{#1}\fi
\ifx \bseriesno  \undefined \def \bseriesno#1{#1}\fi
\ifx \blocation  \undefined \def \blocation#1{#1}\fi
\ifx \bsertitle  \undefined \def \bsertitle#1{#1}\fi
\ifx \bsnm \undefined \def \bsnm#1{#1}\fi
\ifx \bsuffix \undefined \def \bsuffix#1{#1}\fi
\ifx \bparticle \undefined \def \bparticle#1{#1}\fi
\ifx \barticle \undefined \def \barticle#1{#1}\fi
\bibcommenthead
\ifx \bconfdate \undefined \def \bconfdate #1{#1}\fi
\ifx \botherref \undefined \def \botherref #1{#1}\fi
\ifx \url \undefined \def \url#1{\textsf{#1}}\fi
\ifx \bchapter \undefined \def \bchapter#1{#1}\fi
\ifx \bbook \undefined \def \bbook#1{#1}\fi
\ifx \bcomment \undefined \def \bcomment#1{#1}\fi
\ifx \oauthor \undefined \def \oauthor#1{#1}\fi
\ifx \citeauthoryear \undefined \def \citeauthoryear#1{#1}\fi
\ifx \endbibitem  \undefined \def \endbibitem {}\fi
\ifx \bconflocation  \undefined \def \bconflocation#1{#1}\fi
\ifx \arxivurl  \undefined \def \arxivurl#1{\textsf{#1}}\fi
\csname PreBibitemsHook\endcsname

\bibitem[\protect\citeauthoryear{DellaPosta et~al.}{2015}]{dellaposta2015liberals}
\begin{barticle}
\bauthor{\bsnm{DellaPosta}, \binits{D.}},
\bauthor{\bsnm{Shi}, \binits{Y.}},
\bauthor{\bsnm{Macy}, \binits{M.}}:
\batitle{Why do liberals drink lattes?}
\bjtitle{American Journal of Sociology}
\bvolume{120}(\bissue{5}),
\bfpage{1473}--\blpage{1511}
(\byear{2015})
\end{barticle}
\endbibitem

\bibitem[\protect\citeauthoryear{Rokkan}{1967}]{rokkan1967geography}
\begin{barticle}
\bauthor{\bsnm{Rokkan}, \binits{S.}}:
\batitle{Geography, religion, and social class: Crosscutting cleavages in norwegian politics}.
\bjtitle{Party systems and voter alignments}
\bvolume{367},
\bfpage{379}--\blpage{86}
(\byear{1967})
\end{barticle}
\endbibitem

\bibitem[\protect\citeauthoryear{Blau and Schwartz}{1984}]{blau1984crosscutting}
\begin{botherref}
\oauthor{\bsnm{Blau}, \binits{P.M.}},
\oauthor{\bsnm{Schwartz}, \binits{J.E.}}:
Crosscutting social circles: Testing a macrostructural theory of intergroup relations
(1984)
\end{botherref}
\endbibitem

\bibitem[\protect\citeauthoryear{Mason}{2016}]{mason2016cross}
\begin{barticle}
\bauthor{\bsnm{Mason}, \binits{L.}}:
\batitle{A cross-cutting calm: How social sorting drives affective polarization}.
\bjtitle{Public Opinion Quarterly}
\bvolume{80}(\bissue{S1}),
\bfpage{351}--\blpage{377}
(\byear{2016})
\end{barticle}
\endbibitem

\bibitem[\protect\citeauthoryear{Finkel et~al.}{2020}]{finkel2020political}
\begin{barticle}
\bauthor{\bsnm{Finkel}, \binits{E.J.}},
\bauthor{\bsnm{Bail}, \binits{C.A.}},
\bauthor{\bsnm{Cikara}, \binits{M.}},
\bauthor{\bsnm{Ditto}, \binits{P.H.}},
\bauthor{\bsnm{Iyengar}, \binits{S.}},
\bauthor{\bsnm{Klar}, \binits{S.}},
\bauthor{\bsnm{Mason}, \binits{L.}},
\bauthor{\bsnm{McGrath}, \binits{M.C.}},
\bauthor{\bsnm{Nyhan}, \binits{B.}},
\bauthor{\bsnm{Rand}, \binits{D.G.}}, \betal:
\batitle{Political sectarianism in america}.
\bjtitle{Science}
\bvolume{370}(\bissue{6516}),
\bfpage{533}--\blpage{536}
(\byear{2020})
\end{barticle}
\endbibitem

\bibitem[\protect\citeauthoryear{Lipset et~al.}{1967}]{lipset1967party}
\begin{bbook}
\bauthor{\bsnm{Lipset}, \binits{S.M.}},
\bauthor{\bsnm{Lipset}, \binits{S.M.}},
\bauthor{\bsnm{Rokkan}, \binits{S.}}:
\bbtitle{Party Systems and Voter Alignments: Cross-national Perspectives}
vol. \bseriesno{7}.
\bpublisher{New York: Free Press}, \blocation{???}
(\byear{1967})
\end{bbook}
\endbibitem

\bibitem[\protect\citeauthoryear{Franklin}{1992}]{franklin1992decline}
\begin{botherref}
\oauthor{\bsnm{Franklin}, \binits{M.N.}}:
The decline of cleavage politics.
Electoral change: Responses to evolving social and attitudinal structures in Western countries,
383--405
(1992)
\end{botherref}
\endbibitem

\bibitem[\protect\citeauthoryear{Kriesi et~al.}{2008}]{kriesi2008west}
\begin{bbook}
\bauthor{\bsnm{Kriesi}, \binits{H.}},
\bauthor{\bsnm{Grande}, \binits{E.}},
\bauthor{\bsnm{Lachat}, \binits{R.}},
\bauthor{\bsnm{Dolezal}, \binits{M.}},
\bauthor{\bsnm{Bornschier}, \binits{S.}},
\bauthor{\bsnm{Frey}, \binits{T.}}:
\bbtitle{West European Politics in the Age of Globalization}.
\bpublisher{Cambridge University Press}, \blocation{???}
(\byear{2008})
\end{bbook}
\endbibitem

\bibitem[\protect\citeauthoryear{Ford and Jennings}{2020}]{ford2020changing}
\begin{barticle}
\bauthor{\bsnm{Ford}, \binits{R.}},
\bauthor{\bsnm{Jennings}, \binits{W.}}:
\batitle{The changing cleavage politics of western europe}.
\bjtitle{Annual review of political science}
\bvolume{23},
\bfpage{295}--\blpage{314}
(\byear{2020})
\end{barticle}
\endbibitem

\bibitem[\protect\citeauthoryear{Hooghe and Marks}{2018}]{hooghe2018cleavage}
\begin{barticle}
\bauthor{\bsnm{Hooghe}, \binits{L.}},
\bauthor{\bsnm{Marks}, \binits{G.}}:
\batitle{Cleavage theory meets europe’s crises: Lipset, rokkan, and the transnational cleavage}.
\bjtitle{Journal of European public policy}
\bvolume{25}(\bissue{1}),
\bfpage{109}--\blpage{135}
(\byear{2018})
\end{barticle}
\endbibitem

\bibitem[\protect\citeauthoryear{Bartolini and Mair}{2007}]{bartolini2007identity}
\begin{bbook}
\bauthor{\bsnm{Bartolini}, \binits{S.}},
\bauthor{\bsnm{Mair}, \binits{P.}}:
\bbtitle{Identity, Competition and Electoral Availability: the Stabilisation of European Electorates 1885-1985}.
\bpublisher{ECPR Press}, \blocation{???}
(\byear{2007})
\end{bbook}
\endbibitem

\bibitem[\protect\citeauthoryear{Goldberg}{2020}]{goldberg2020evolution}
\begin{barticle}
\bauthor{\bsnm{Goldberg}, \binits{A.C.}}:
\batitle{The evolution of cleavage voting in four western countries: Structural, behavioural or political dealignment?}
\bjtitle{European Journal of Political Research}
\bvolume{59}(\bissue{1}),
\bfpage{68}--\blpage{90}
(\byear{2020})
\end{barticle}
\endbibitem

\bibitem[\protect\citeauthoryear{Guerra et~al.}{2013}]{guerra2013measure}
\begin{bchapter}
\bauthor{\bsnm{Guerra}, \binits{P.C.}},
\bauthor{\bsnm{Meira~Jr}, \binits{W.}},
\bauthor{\bsnm{Cardie}, \binits{C.}},
\bauthor{\bsnm{Kleinberg}, \binits{R.}}:
\bctitle{A measure of polarization on social media networks based on community boundaries}.
In: \bbtitle{Seventh International AAAI Conference on Weblogs and Social Media}
(\byear{2013})
\end{bchapter}
\endbibitem

\bibitem[\protect\citeauthoryear{Keuchenius et~al.}{2021}]{keuchenius2021important}
\begin{barticle}
\bauthor{\bsnm{Keuchenius}, \binits{A.}},
\bauthor{\bsnm{T{\"o}rnberg}, \binits{P.}},
\bauthor{\bsnm{Uitermark}, \binits{J.}}:
\batitle{Why it is important to consider negative ties when studying polarized debates: A signed network analysis of a dutch cultural controversy on twitter}.
\bjtitle{PloS one}
\bvolume{16}(\bissue{8}),
\bfpage{0256696}
(\byear{2021})
\end{barticle}
\endbibitem

\bibitem[\protect\citeauthoryear{Barber{\'a} et~al.}{2015}]{barbera2015tweeting}
\begin{barticle}
\bauthor{\bsnm{Barber{\'a}}, \binits{P.}},
\bauthor{\bsnm{Jost}, \binits{J.T.}},
\bauthor{\bsnm{Nagler}, \binits{J.}},
\bauthor{\bsnm{Tucker}, \binits{J.A.}},
\bauthor{\bsnm{Bonneau}, \binits{R.}}:
\batitle{Tweeting from left to right: Is online political communication more than an echo chamber?}
\bjtitle{Psychological science}
\bvolume{26}(\bissue{10}),
\bfpage{1531}--\blpage{1542}
(\byear{2015})
\end{barticle}
\endbibitem

\bibitem[\protect\citeauthoryear{Heider}{1958}]{heiderPsychologyInterpersonalRelations1958}
\begin{bbook}
\bauthor{\bsnm{Heider}, \binits{F.}}:
\bbtitle{The Psychology of Interpersonal Relations.}
\bpublisher{{John Wiley \& Sons Inc}},
\blocation{{Hoboken}}
(\byear{1958}).
\doiurl{10.1037/10628-000}
\end{bbook}
\endbibitem

\bibitem[\protect\citeauthoryear{Cartwright and Harary}{1956}]{cartwright1956structural}
\begin{barticle}
\bauthor{\bsnm{Cartwright}, \binits{D.}},
\bauthor{\bsnm{Harary}, \binits{F.}}:
\batitle{Structural balance: a generalization of heider's theory.}
\bjtitle{Psychological review}
\bvolume{63}(\bissue{5}),
\bfpage{277}
(\byear{1956})
\end{barticle}
\endbibitem

\bibitem[\protect\citeauthoryear{Leskovec et~al.}{2010}]{leskovec2010signed}
\begin{bchapter}
\bauthor{\bsnm{Leskovec}, \binits{J.}},
\bauthor{\bsnm{Huttenlocher}, \binits{D.}},
\bauthor{\bsnm{Kleinberg}, \binits{J.}}:
\bctitle{Signed networks in social media}.
In: \bbtitle{Proceedings of the SIGCHI Conference on Human Factors in Computing Systems},
pp. \bfpage{1361}--\blpage{1370}
(\byear{2010})
\end{bchapter}
\endbibitem

\bibitem[\protect\citeauthoryear{Estrada and Benzi}{2014}]{estrada2014walk}
\begin{barticle}
\bauthor{\bsnm{Estrada}, \binits{E.}},
\bauthor{\bsnm{Benzi}, \binits{M.}}:
\batitle{Walk-based measure of balance in signed networks: Detecting lack of balance in social networks}.
\bjtitle{Physical Review E}
\bvolume{90}(\bissue{4}),
\bfpage{042802}
(\byear{2014})
\end{barticle}
\endbibitem

\bibitem[\protect\citeauthoryear{Aref and Wilson}{2018}]{aref2018measuring}
\begin{barticle}
\bauthor{\bsnm{Aref}, \binits{S.}},
\bauthor{\bsnm{Wilson}, \binits{M.C.}}:
\batitle{Measuring partial balance in signed networks}.
\bjtitle{Journal of Complex Networks}
\bvolume{6}(\bissue{4}),
\bfpage{566}--\blpage{595}
(\byear{2018})
\end{barticle}
\endbibitem

\bibitem[\protect\citeauthoryear{Aref et~al.}{2020}]{aref2020multilevel}
\begin{barticle}
\bauthor{\bsnm{Aref}, \binits{S.}},
\bauthor{\bsnm{Dinh}, \binits{L.}},
\bauthor{\bsnm{Rezapour}, \binits{R.}},
\bauthor{\bsnm{Diesner}, \binits{J.}}:
\batitle{Multilevel structural evaluation of signed directed social networks based on balance theory}.
\bjtitle{Scientific reports}
\bvolume{10}(\bissue{1}),
\bfpage{1}--\blpage{12}
(\byear{2020})
\end{barticle}
\endbibitem

\bibitem[\protect\citeauthoryear{Andres et~al.}{2022}]{andres2022reconstructing}
\begin{botherref}
\oauthor{\bsnm{Andres}, \binits{G.}},
\oauthor{\bsnm{Casiraghi}, \binits{G.}},
\oauthor{\bsnm{Vaccario}, \binits{G.}},
\oauthor{\bsnm{Schweitzer}, \binits{F.}}:
Reconstructing signed relations from interaction data.
arXiv preprint arXiv:2209.03219
(2022)
\end{botherref}
\endbibitem

\bibitem[\protect\citeauthoryear{Garcia and Tanase}{2013}]{garcia2013measuring}
\begin{barticle}
\bauthor{\bsnm{Garcia}, \binits{D.}},
\bauthor{\bsnm{Tanase}, \binits{D.}}:
\batitle{Measuring cultural dynamics through the eurovision song contest}.
\bjtitle{Advances in Complex Systems}
\bvolume{16}(\bissue{08}),
\bfpage{1350037}
(\byear{2013})
\end{barticle}
\endbibitem

\bibitem[\protect\citeauthoryear{Neal}{2014}]{neal2014backbone}
\begin{barticle}
\bauthor{\bsnm{Neal}, \binits{Z.}}:
\batitle{The backbone of bipartite projections: Inferring relationships from co-authorship, co-sponsorship, co-attendance and other co-behaviors}.
\bjtitle{Social Networks}
\bvolume{39},
\bfpage{84}--\blpage{97}
(\byear{2014})
\end{barticle}
\endbibitem

\bibitem[\protect\citeauthoryear{Tufekci}{2014}]{tufekci2014big}
\begin{bchapter}
\bauthor{\bsnm{Tufekci}, \binits{Z.}}:
\bctitle{Big questions for social media big data: Representativeness, validity and other methodological pitfalls}.
In: \bbtitle{Proceedings of the International AAAI Conference on Web and Social Media},
vol. \bseriesno{8},
pp. \bfpage{505}--\blpage{514}
(\byear{2014})
\end{bchapter}
\endbibitem

\bibitem[\protect\citeauthoryear{Doreian and Mrvar}{2015}]{doreian2015structural}
\begin{barticle}
\bauthor{\bsnm{Doreian}, \binits{P.}},
\bauthor{\bsnm{Mrvar}, \binits{A.}}:
\batitle{Structural balance and signed international relations}.
\bjtitle{Journal of Social Structure}
\bvolume{16}(\bissue{1}),
\bfpage{1}--\blpage{49}
(\byear{2015})
\end{barticle}
\endbibitem

\bibitem[\protect\citeauthoryear{Maoz et~al.}{2007}]{maoz2007enemy}
\begin{barticle}
\bauthor{\bsnm{Maoz}, \binits{Z.}},
\bauthor{\bsnm{Terris}, \binits{L.G.}},
\bauthor{\bsnm{Kuperman}, \binits{R.D.}},
\bauthor{\bsnm{Talmud}, \binits{I.}}:
\batitle{What is the enemy of my enemy? causes and consequences of imbalanced international relations, 1816--2001}.
\bjtitle{The Journal of Politics}
\bvolume{69}(\bissue{1}),
\bfpage{100}--\blpage{115}
(\byear{2007})
\end{barticle}
\endbibitem

\bibitem[\protect\citeauthoryear{Diaz-Diaz et~al.}{2023}]{diaz2023network}
\begin{botherref}
\oauthor{\bsnm{Diaz-Diaz}, \binits{F.}},
\oauthor{\bsnm{Bartesaghi}, \binits{P.}},
\oauthor{\bsnm{Estrada}, \binits{E.}}:
Network theory meets history. local balance in global international relations.
arXiv preprint arXiv:2303.03774
(2023)
\end{botherref}
\endbibitem

\bibitem[\protect\citeauthoryear{Estrada}{2019}]{estrada2019rethinking}
\begin{barticle}
\bauthor{\bsnm{Estrada}, \binits{E.}}:
\batitle{Rethinking structural balance in signed social networks}.
\bjtitle{Discrete Applied Mathematics}
\bvolume{268},
\bfpage{70}--\blpage{90}
(\byear{2019})
\end{barticle}
\endbibitem

\bibitem[\protect\citeauthoryear{Aref and Neal}{2021}]{aref2021identifying}
\begin{barticle}
\bauthor{\bsnm{Aref}, \binits{S.}},
\bauthor{\bsnm{Neal}, \binits{Z.P.}}:
\batitle{Identifying hidden coalitions in the us house of representatives by optimally partitioning signed networks based on generalized balance}.
\bjtitle{Scientific reports}
\bvolume{11}(\bissue{1}),
\bfpage{1}--\blpage{9}
(\byear{2021})
\end{barticle}
\endbibitem

\bibitem[\protect\citeauthoryear{Guha et~al.}{2004}]{guha2004propagation}
\begin{bchapter}
\bauthor{\bsnm{Guha}, \binits{R.}},
\bauthor{\bsnm{Kumar}, \binits{R.}},
\bauthor{\bsnm{Raghavan}, \binits{P.}},
\bauthor{\bsnm{Tomkins}, \binits{A.}}:
\bctitle{Propagation of trust and distrust}.
In: \bbtitle{Proceedings of the 13th International Conference on World Wide Web},
pp. \bfpage{403}--\blpage{412}
(\byear{2004})
\end{bchapter}
\endbibitem

\bibitem[\protect\citeauthoryear{Kunegis et~al.}{2009}]{kunegis2009slashdot}
\begin{bchapter}
\bauthor{\bsnm{Kunegis}, \binits{J.}},
\bauthor{\bsnm{Lommatzsch}, \binits{A.}},
\bauthor{\bsnm{Bauckhage}, \binits{C.}}:
\bctitle{The slashdot zoo: mining a social network with negative edges}.
In: \bbtitle{Proceedings of the 18th International Conference on World Wide Web},
pp. \bfpage{741}--\blpage{750}
(\byear{2009})
\end{bchapter}
\endbibitem

\bibitem[\protect\citeauthoryear{West et~al.}{2014}]{west2014exploiting}
\begin{barticle}
\bauthor{\bsnm{West}, \binits{R.}},
\bauthor{\bsnm{Paskov}, \binits{H.S.}},
\bauthor{\bsnm{Leskovec}, \binits{J.}},
\bauthor{\bsnm{Potts}, \binits{C.}}:
\batitle{Exploiting social network structure for person-to-person sentiment analysis}.
\bjtitle{Transactions of the Association for Computational Linguistics}
\bvolume{2},
\bfpage{297}--\blpage{310}
(\byear{2014})
\end{barticle}
\endbibitem

\bibitem[\protect\citeauthoryear{Maniu et~al.}{2011}]{maniuBuildingSignedNetwork2011}
\begin{bchapter}
\bauthor{\bsnm{Maniu}, \binits{S.}},
\bauthor{\bsnm{Cautis}, \binits{B.}},
\bauthor{\bsnm{Abdessalem}, \binits{T.}}:
\bctitle{Building a signed network from interactions in {{Wikipedia}}}.
In: \bbtitle{Databases and {{Social}} {{Networks}} on - {{DBSocial}} '11},
pp. \bfpage{19}--\blpage{24}.
\bpublisher{{ACM Press}},
\blocation{{Athens, Greece}}
(\byear{2011}).
\doiurl{10.1145/1996413.1996417}
\end{bchapter}
\endbibitem

\bibitem[\protect\citeauthoryear{Pougu{\'e}-Biyong et~al.}{2021}]{pougue2021debagreement}
\begin{bchapter}
\bauthor{\bsnm{Pougu{\'e}-Biyong}, \binits{J.}},
\bauthor{\bsnm{Semenova}, \binits{V.}},
\bauthor{\bsnm{Matton}, \binits{A.}},
\bauthor{\bsnm{Han}, \binits{R.}},
\bauthor{\bsnm{Kim}, \binits{A.}},
\bauthor{\bsnm{Lambiotte}, \binits{R.}},
\bauthor{\bsnm{Farmer}, \binits{D.}}:
\bctitle{Debagreement: A comment-reply dataset for (dis) agreement detection in online debates}.
In: \bbtitle{Thirty-fifth Conference on Neural Information Processing Systems Datasets and Benchmarks Track (Round 2)}
(\byear{2021})
\end{bchapter}
\endbibitem

\bibitem[\protect\citeauthoryear{Pougu{\'e}-Biyong et~al.}{2022}]{pougue2022learning}
\begin{botherref}
\oauthor{\bsnm{Pougu{\'e}-Biyong}, \binits{J.}},
\oauthor{\bsnm{Gupta}, \binits{A.}},
\oauthor{\bsnm{Haghighi}, \binits{A.}},
\oauthor{\bsnm{El-Kishky}, \binits{A.}}:
Learning stance embeddings from signed social graphs.
arXiv preprint arXiv:2201.11675
(2022)
\end{botherref}
\endbibitem

\bibitem[\protect\citeauthoryear{Pr{\"o}llochs}{2022}]{prollochs2022community}
\begin{bchapter}
\bauthor{\bsnm{Pr{\"o}llochs}, \binits{N.}}:
\bctitle{Community-based fact-checking on twitter’s birdwatch platform}.
In: \bbtitle{Proceedings of the International AAAI Conference on Web and Social Media},
vol. \bseriesno{16},
pp. \bfpage{794}--\blpage{805}
(\byear{2022})
\end{bchapter}
\endbibitem

\bibitem[\protect\citeauthoryear{Saeed et~al.}{2022}]{saeed2022crowdsourced}
\begin{bchapter}
\bauthor{\bsnm{Saeed}, \binits{M.}},
\bauthor{\bsnm{Traub}, \binits{N.}},
\bauthor{\bsnm{Nicolas}, \binits{M.}},
\bauthor{\bsnm{Demartini}, \binits{G.}},
\bauthor{\bsnm{Papotti}, \binits{P.}}:
\bctitle{Crowdsourced fact-checking at twitter: How does the crowd compare with experts?}
In: \bbtitle{Proceedings of the 31st ACM International Conference on Information \& Knowledge Management},
pp. \bfpage{1736}--\blpage{1746}
(\byear{2022})
\end{bchapter}
\endbibitem

\bibitem[\protect\citeauthoryear{Drolsbach and Pr{\"o}llochs}{2023}]{drolsbach2023believability}
\begin{bchapter}
\bauthor{\bsnm{Drolsbach}, \binits{C.P.}},
\bauthor{\bsnm{Pr{\"o}llochs}, \binits{N.}}:
\bctitle{Believability and harmfulness shape the virality of misleading social media posts}.
In: \bbtitle{Proceedings of the ACM Web Conference 2023},
pp. \bfpage{4172}--\blpage{4177}
(\byear{2023})
\end{bchapter}
\endbibitem

\bibitem[\protect\citeauthoryear{Wojcik et~al.}{2022}]{wojcik2022birdwatch}
\begin{botherref}
\oauthor{\bsnm{Wojcik}, \binits{S.}},
\oauthor{\bsnm{Hilgard}, \binits{S.}},
\oauthor{\bsnm{Judd}, \binits{N.}},
\oauthor{\bsnm{Mocanu}, \binits{D.}},
\oauthor{\bsnm{Ragain}, \binits{S.}},
\oauthor{\bsnm{Hunzaker}, \binits{M.}},
\oauthor{\bsnm{Coleman}, \binits{K.}},
\oauthor{\bsnm{Baxter}, \binits{J.}}:
Birdwatch: Crowd wisdom and bridging algorithms can inform understanding and reduce the spread of misinformation.
arXiv preprint arXiv:2210.15723
(2022)
\end{botherref}
\endbibitem

\bibitem[\protect\citeauthoryear{Drolsbach and Pr{\"o}llochs}{2022}]{drolsbach2022diffusion}
\begin{botherref}
\oauthor{\bsnm{Drolsbach}, \binits{C.}},
\oauthor{\bsnm{Pr{\"o}llochs}, \binits{N.}}:
Diffusion of community fact-checked misinformation on twitter.
arXiv preprint arXiv:2205.13673
(2022)
\end{botherref}
\endbibitem

\bibitem[\protect\citeauthoryear{Allen et~al.}{2022}]{allen2022birds}
\begin{bchapter}
\bauthor{\bsnm{Allen}, \binits{J.}},
\bauthor{\bsnm{Martel}, \binits{C.}},
\bauthor{\bsnm{Rand}, \binits{D.G.}}:
\bctitle{Birds of a feather don’t fact-check each other: Partisanship and the evaluation of news in twitter’s birdwatch crowdsourced fact-checking program}.
In: \bbtitle{Proceedings of the 2022 CHI Conference on Human Factors in Computing Systems},
pp. \bfpage{1}--\blpage{19}
(\byear{2022})
\end{bchapter}
\endbibitem

\bibitem[\protect\citeauthoryear{Niederkrotenthaler et~al.}{2022}]{niederkrotenthaler2022mental}
\begin{barticle}
\bauthor{\bsnm{Niederkrotenthaler}, \binits{T.}},
\bauthor{\bsnm{Laido}, \binits{Z.}},
\bauthor{\bsnm{Kirchner}, \binits{S.}},
\bauthor{\bsnm{Braun}, \binits{M.}},
\bauthor{\bsnm{Metzler}, \binits{H.}},
\bauthor{\bsnm{Waldh{\"o}r}, \binits{T.}},
\bauthor{\bsnm{Strauss}, \binits{M.}},
\bauthor{\bsnm{Garcia}, \binits{D.}},
\bauthor{\bsnm{Till}, \binits{B.}}:
\batitle{Mental health over nine months during the sars-cov2 pandemic: Representative cross-sectional survey in twelve waves between april and december 2020 in austria}.
\bjtitle{Journal of affective disorders}
\bvolume{296},
\bfpage{49}--\blpage{58}
(\byear{2022})
\end{barticle}
\endbibitem

\bibitem[\protect\citeauthoryear{Aref et~al.}{2020}]{aref2020modeling}
\begin{barticle}
\bauthor{\bsnm{Aref}, \binits{S.}},
\bauthor{\bsnm{Mason}, \binits{A.J.}},
\bauthor{\bsnm{Wilson}, \binits{M.C.}}:
\batitle{A modeling and computational study of the frustration index in signed networks}.
\bjtitle{Networks}
\bvolume{75}(\bissue{1}),
\bfpage{95}--\blpage{110}
(\byear{2020})
\end{barticle}
\endbibitem

\bibitem[\protect\citeauthoryear{Doreian and Mrvar}{2009}]{doreian2009partitioning}
\begin{barticle}
\bauthor{\bsnm{Doreian}, \binits{P.}},
\bauthor{\bsnm{Mrvar}, \binits{A.}}:
\batitle{Partitioning signed social networks}.
\bjtitle{Social Networks}
\bvolume{31}(\bissue{1}),
\bfpage{1}--\blpage{11}
(\byear{2009})
\end{barticle}
\endbibitem

\bibitem[\protect\citeauthoryear{Schoch}{2020}]{signnet}
\begin{botherref}
\oauthor{\bsnm{Schoch}, \binits{D.}}:
Signnet: An R Package to Analyze Signed Networks.
(2020).
\url{https://github.com/schochastics/signnet}
\end{botherref}
\endbibitem

\bibitem[\protect\citeauthoryear{Davis}{1967}]{davis1967clustering}
\begin{barticle}
\bauthor{\bsnm{Davis}, \binits{J.A.}}:
\batitle{Clustering and structural balance in graphs}.
\bjtitle{Human relations}
\bvolume{20}(\bissue{2}),
\bfpage{181}--\blpage{187}
(\byear{1967})
\end{barticle}
\endbibitem

\bibitem[\protect\citeauthoryear{Vijaymeena and Kavitha}{2016}]{vijaymeena2016survey}
\begin{barticle}
\bauthor{\bsnm{Vijaymeena}, \binits{M.}},
\bauthor{\bsnm{Kavitha}, \binits{K.}}:
\batitle{A survey on similarity measures in text mining}.
\bjtitle{Machine Learning and Applications: An International Journal}
\bvolume{3}(\bissue{2}),
\bfpage{19}--\blpage{28}
(\byear{2016})
\end{barticle}
\endbibitem

\bibitem[\protect\citeauthoryear{Gallagher et~al.}{2021}]{gallagher2021generalized}
\begin{barticle}
\bauthor{\bsnm{Gallagher}, \binits{R.J.}},
\bauthor{\bsnm{Frank}, \binits{M.R.}},
\bauthor{\bsnm{Mitchell}, \binits{L.}},
\bauthor{\bsnm{Schwartz}, \binits{A.J.}},
\bauthor{\bsnm{Reagan}, \binits{A.J.}},
\bauthor{\bsnm{Danforth}, \binits{C.M.}},
\bauthor{\bsnm{Dodds}, \binits{P.S.}}:
\batitle{Generalized word shift graphs: a method for visualizing and explaining pairwise comparisons between texts}.
\bjtitle{EPJ Data Science}
\bvolume{10}(\bissue{1}),
\bfpage{4}
(\byear{2021})
\end{barticle}
\endbibitem

\bibitem[\protect\citeauthoryear{Wikipedia}{2023}]{dewiki:236477099}
\begin{botherref}
\oauthor{\bsnm{Wikipedia}}:
Querfront --- Wikipedia{,} die freie Enzyklopädie.
[Online; Stand 12. September 2023]
(2023).
\url{https://de.wikipedia.org/w/index.php?title=Querfront&oldid=236477099}
\end{botherref}
\endbibitem

\bibitem[\protect\citeauthoryear{}{2023a}]{2015EuropeanMigrant2023}
\begin{botherref}
2015 {European} migrant crisis.
Page Version ID: 1159102024
(2023).
\url{https://en.wikipedia.org/w/index.php?title=2015_European_migrant_crisis&oldid=1159102024}
Accessed 2023-06-14
\end{botherref}
\endbibitem

\bibitem[\protect\citeauthoryear{}{2023b}]{201516New2023}
\begin{botherref}
2015–16 {New} {Year}'s {Eve} sexual assaults in {Germany}.
Page Version ID: 1159999875
(2023).
\url{https://en.wikipedia.org/w/index.php?title=2015%E2%80%9316_New_Year%27s_Eve_sexual_assaults_in_Germany&oldid=1159999875}
Accessed 2023-06-14
\end{botherref}
\endbibitem

\bibitem[\protect\citeauthoryear{Garcia et~al.}{2015}]{garcia2015ideological}
\begin{barticle}
\bauthor{\bsnm{Garcia}, \binits{D.}},
\bauthor{\bsnm{Abisheva}, \binits{A.}},
\bauthor{\bsnm{Schweighofer}, \binits{S.}},
\bauthor{\bsnm{Serd{\"u}lt}, \binits{U.}},
\bauthor{\bsnm{Schweitzer}, \binits{F.}}:
\batitle{Ideological and temporal components of network polarization in online political participatory media}.
\bjtitle{Policy \& internet}
\bvolume{7}(\bissue{1}),
\bfpage{46}--\blpage{79}
(\byear{2015})
\end{barticle}
\endbibitem

\bibitem[\protect\citeauthoryear{Baldassarri and Gelman}{2008}]{baldassarri2008partisans}
\begin{barticle}
\bauthor{\bsnm{Baldassarri}, \binits{D.}},
\bauthor{\bsnm{Gelman}, \binits{A.}}:
\batitle{Partisans without constraint: Political polarization and trends in american public opinion}.
\bjtitle{American Journal of Sociology}
\bvolume{114}(\bissue{2}),
\bfpage{408}--\blpage{446}
(\byear{2008})
\end{barticle}
\endbibitem

\end{thebibliography}


\end{document}


\section{Supplementary Materials}\label{SI}

This pdf file includes: 

\begin{itemize}
    \item Supplementary text
    \item Supplementary figures 1-9
    \item Supplementary tables 1-2
\end{itemize}

\subsection{S1. Supplementary text. Derstandard Contextualization}\label{derst_context}

\begin{figure}[!htbp]
    \centering
    \includegraphics[width=\linewidth]{figures_SI/votes_example.png}
    \caption{\textbf{Example of one DerStandard forum post, showing votes.} Each posting in the forum can be up (green) or down votes by other registered user. The bar on the left top of the posting shows the sum for both types of votes. By clicking on that bar, we open a menu that contains the user names of voters and the type of vote cast. (User names have been blurred by the authors in this example.)}
    \label{fig:votes_screenshot}
\end{figure}

\textbf{Context: 2015-2016 changes in Austria}

It's striking that between 2015 and 2016 a pronounced increase occurs in our Alignment metric for political topics in DerStandard. During that time, media discourse was dominated by the events in Cologne (and other German cities) during the New Year's Eve 15/16 celebrations. A substantial number of women was reporting sexual assault by groups in public, an unusual criminal offense in Germany (for a timeline of events and contextualisation see \cite{boulila2017cologne} and \href{https://en.wikipedia.org/wiki/2015%E2%80%9316_New_Year%27s_Eve_sexual_assaults_in_Germany}{Wikipedia page}). The political discussion starting with those events led to a pronounced shift in public opinion, summarized by the influential German newspaper "Der Spiegel" as \href{https://www.spiegel.de/international/germany/is-there-truth-to-refugee-sex-offense-reports-a-1186734.html}{such}: "The night brought an end to the sense of euphoria that had accompanied the welcoming of hundreds of thousands of refugees into the country earlier that year".

\subsubsection*{S2. Supplementary text. Assessment of approximated results}

To assess the robustness of the results, given that we use an approximated measure for some of the datasets, we run the partitioning algorithm $200$ times for each network and keep the lowest value of frustrated edges and its respective partition. We provide three robustness checks: 
(i) To assess the number of iterations of the algorithm, we check if we can find the optimal solution within $it = \frac{1}{2}200$ iterations, and if not we see how different would the final result change. We find that the number of frustrated edges of the optimal solution found in half of the iterations differs less than $1\%$ with the one found in $200$ iterations.
(ii) We compare the similarities between the partitions within the 3 best solutions found: in the three cases we find almost-identical sizes for the obtained groups, with a Szymkiewicz–Simpson overlap coefficient of $0.79$.
(iii) We provide a comparison analysis between the exact solutions and approximated solutions when using a rolling window for the BW1 dataset, see Figure~\ref{fig:comparison}. The groups in the exact and approximated partitions have a Szymkiewicz–Simpson overlap coefficient of $0.8$. The Pearson correlation coefficient between the obtained time series is $0.868$.

Tables \ref{t:bw_results} and \ref{t:der_results} show the detailed resulting partitions from the exact or approximated methods.

\begin{landscape}
\begin{table}[]
\centering
\resizebox{\linewidth}{!}{%
\begin{tabular}{c|c|c|c|c|c|c|c|c|c|c|c}
 &
  \textbf{Method} &
  \textbf{\begin{tabular}[c]{@{}c@{}}\ $K^*$\end{tabular}} &
  \textbf{\begin{tabular}[c]{@{}c@{}}Ratio \\ Size\\ Groups\end{tabular}} &
  \textbf{\begin{tabular}[c]{@{}c@{}}Ratio \\ Internal/\\ External\end{tabular}} &
  \textbf{\begin{tabular}[c]{@{}c@{}}\%\\  Frus \\ Edges\end{tabular}} &
  \textbf{\begin{tabular}[c]{@{}c@{}}Coh/\\ Norm. Coh\end{tabular}} &
  \textbf{\begin{tabular}[c]{@{}c@{}}Coh \\ Bootstrapping \\ Interval \end{tabular}} &
  \textbf{\begin{tabular}[c]{@{}c@{}}Div/\\ Norm. Div\end{tabular}} &
  \textbf{\begin{tabular}[c]{@{}c@{}}Div \\ Bootstrapping \\ Interval \end{tabular}} &
  \textbf{$SAI_G$} &
  \textbf{\begin{tabular}[c]{@{}c@{}} $SAI_G$ 95\% confidence \\ intervals \end{tabular}} \\ \hline
\textbf{\begin{tabular}[c]{@{}c@{}}BW1\end{tabular}} &
  Exact &
  2 &
  65/35 &
  67/33 &
  14\% &
  0.929 / 0.201 &
  0.004 &
  0.72 / 0.43 &
  0.01 &
  0.668 &
  0.6642-0.6722 \\ \hline
\textbf{\begin{tabular}[c]{@{}c@{}}BW1\end{tabular}} &
  Approx&
  2 &
  71/29  &
  66/34 &
  18\% &
  0.901 / 0.182 &
  0.005 &
  0.64 / 0.36 &
  0.01 &
  0.563 &
  0.5579-5686 \\ \hline
\textbf{\begin{tabular}[c]{@{}c@{}}BW2\end{tabular}} &
  Approx. &
  2 &
  73/27 &
  62/38 &
  25\% &
  0.801 / 0.181 &
  0.002 &
  0.674 / 0.294 &
  0.003 &
  0.475 &
  0.4738-0.4780 
\end{tabular}%
}
\caption{\textbf{Birdwatch Global Results.} Summary of the optimal partition results for the two datasets. To evaluate the difference in the use of the methods, for BW1, we show results obtained with both the exact and approximated method. $SAI_G$ confidence intervals are obtained from running 10,000 instances of the null model. Divisiveness and Cohesiveness resample uncertainties are obtained by bootstrapping for 10,000 instances.}
\label{t:bw_results}
\end{table}

\begin{table}[]
\centering
\resizebox{\linewidth}{!}{%
\begin{tabular}{c|c|c|c|c|c|c|c|c|c|c|c}
 &
  \textbf{Method} &
  \textbf{\begin{tabular}[c]{@{}c@{}}\ $K^*$\end{tabular}} &
  \textbf{\begin{tabular}[c]{@{}c@{}}Ratio \\ Size\\ Groups\end{tabular}} &
  \textbf{\begin{tabular}[c]{@{}c@{}}Ratio \\ Internal/\\ External\end{tabular}} &
  \textbf{\begin{tabular}[c]{@{}c@{}}\%\\  Frus \\ Edges\end{tabular}} &
  \textbf{\begin{tabular}[c]{@{}c@{}}Coh/\\ Norm. Coh\end{tabular}} &
  \textbf{\begin{tabular}[c]{@{}c@{}}Coh \\ Bootstrapping \\ Interval\end{tabular}} &
  \textbf{\begin{tabular}[c]{@{}c@{}}Div/\\ Norm. Div\end{tabular}} &
  \textbf{\begin{tabular}[c]{@{}c@{}}Div \\ Bootstrapping \\ Interval\end{tabular}} &
  \textbf{$SAI_G$} &
  \textbf{\begin{tabular}[c]{@{}c@{}} $SAI_G$ 95\% confidence \\intervals \end{tabular}} \\ \hline
\textbf{\begin{tabular}[c]{@{}c@{}}DS\end{tabular}} &
  Approx. &
  2 &
  62/37 &
  67/33 &
  29\% &
  0.7006 / 0.1409 &
  0.0002 &
  0.7301 / 0.2899 &
  0.0003 &
  0.396 &
  0.3951-0.3961
\end{tabular}%
}
\caption{\textbf{DerStandard Global Results.} Summary of the optimal partition results for the dataset obtained from DerStandard. $SAI_G$ confidence intervals are obtained from running 1,000 instances of the null model. Divisiveness and Cohesiveness resample uncertainties are obtained by bootstrapping for 10,000 instances.}
\label{t:der_results}
\end{table}
\end{landscape}

\begin{figure}[!htbp]
    \centering
    \includegraphics[width=0.95\linewidth]{figures_SI/SPI_p1_comparison.png}
    \caption{\textbf{Comparison between the timeline results obtained for the approximated and exact methods in the BW1 dataset.} This figure is an analogous of Figure 4 in the main text with different rolling window parameters. It presents the changes in Alignment obtained with the optimal partition of the exact method and the sub-optimal partition obtained through the approximated algorithm with the same data. Even though the approximated results are consistently lower than the exact results, the variations in the two time series are highly correlated. }
    \label{fig:comparison}
\end{figure}

\subsection{S3. Supplementary Figure. Multipartition study}\label{multipartition}

Below we show the distribution of the partitioning algorithm results for the approximated method. This method is of stochastic nature and therefore we run it several times (i.e. 200 instances) and select the partition that yields the minimum number of frustrated edges. We also use these results to select the optimal number of groups, $k^*$, by selecting $k$ with the lowest value of frustrated edges. In Figure \ref{fig:multi} we show how the trend of results seems to increase with $k$, in agreement with the theorem in \cite{doreian2009partitioning} which states that the number of minimum frustrated edges is concave when plotted against $k$. We show the results for the Destandard dataset and for the BW2 dataset, which are the datasets that require the use of the approximated method because of their dimensions. 

\begin{figure}[!htbp]
    \centering
    \includegraphics[width=0.45\linewidth]{figures_SI/multipartition_derst.png}
    \includegraphics[width=0.45\linewidth]{figures_SI/multipartition_part2.png}
    \caption{\textbf{Multipartition study for Derstandard (left) and BW2 (right).} We show the distribution of results for the approximated method for $k=2,3$ and $4$. In a straight line, we mark the best partition result, which we assume to be the closest to the global optima. All other solutions are sub-optimal and therefore local optima. In both datasets $k=2$ is the best number of groups.}
    \label{fig:multi}
\end{figure}

\subsection{S4. Supplementary Figure. Metrics normalization}\label{}

In Fig \ref{fig:null} we show the metrics of Divisiveness and Cohesiveness before normalization for the time series of BW1. We show both the original data metrics and the null model mean. This figure supports the normalization choice of subtracting Antagonism (i.e. proportion of negative interactions) from Divisiveness to obtain a more meaningful signal on the relevance of sign distribution in a specific time window. Cohesiveness, on the other hand, is perfectly correlated with the proportion of positive interactions and thus should be normalized by subtracting such amount. 

\begin{figure}[!htbp]
    \centering
    \includesvg[width=0.95\linewidth]{figures_SI/BW1_cohdiv_null.svg}
    \caption{\textbf{Cohesiveness and Divisiveness of the original data and null model before normalization for the BW1 time series.} We show the metrics of Cohesiveness and Divisiveness for the original data (dotted lines) and for the null model. The null model time series is shown with $95\%$ confidence intervals obtained from the $10,000$ instances of re-shuffled sign distributions. The proportion of negative interactions in each time window is represented in a dashed line and is perfectly correlated to the  mean Divisiveness signal of the null model. Note that it is also inversely correlated to the Cohesiveness of the null model. }
    \label{fig:null}
\end{figure}

\subsection{S5. Supplementary Figure. Antagonism and Alignment in BW1}\label{}

In Fig \ref{fig:aabw1} we show the metrics of Antagonism and Alignment for the time series of BW1. The two time series have a correlation of $0.616$. We find this number to be low enough to identify both metrics as different phenomena and thus to emphasize the importance of considering them separately. Moreover, due to the use of a rolling window for the construction of the time series, this correlation measure also contains auto-correlations, and would otherwise be lower. 

\begin{figure}[!htbp]
    \centering
    \includesvg[width=0.95\linewidth]{figures_SI/BW1_AA.svg}
    \caption{\textbf{Antagonism and Alignment of the BW1 time series.} We see that, while fluctuations are similar for both metrics in some time windows, the correlation between the metrics is low enough to consider them as separate measures that provide different insights. }
    \label{fig:aabw1}
\end{figure}

\subsection{S6. Supplementary Figure. Size effects}\label{size}

In Figure \ref{fig:size_eff} we show the correlation between our Alignment measures for sub-sets of the data and the size of votes (in the case of a temporal rolling window) or votes, posts and articles (in the case of a topic). These coefficients are computed on the data used for the main text figures. As expected, we see there is no direct correlation between the amount of data we consider for each sub-set and the level of Alignment in the network of interactions.

\begin{figure}[!htbp]
    \centering
    \includegraphics[width=0.45\linewidth]{figures_SI/corr_bw1.png}
    \includegraphics[width=0.45\linewidth]{figures_SI/corr_der1.png}
    \includegraphics[width=0.45\linewidth]{figures_SI/corr_der2.png}
    \includegraphics[width=0.45\linewidth]{figures_SI/corr_der3.png}
    \caption{\textbf{Correlation between our Alignment metric and the volume of data of studied subsets.} Scatter plots showing the correlation of the $SAI_R$ measure against the volume of votes for the timeline of BW1 (top left) and the Antagonism-Alignment study for Derstandard (top right). The two lower figures similarly indicate the correlation between the Alginment measures  and the volume of articles and posts obtained for each issue for the Derstandard study.}
    \label{fig:size_eff}
\end{figure}

\subsection{S7. Supplementary Figure. Birdwatch wordshift graphs}\label{wordshift}

We apply a word shifts method in order to contextualize the topics of discussion surrounding the peaks we detect in the timeline obtained for BW1. Word shifts extract which words contribute to a difference between two texts and how they do so. We use the tool \textit{Shifterator} \cite{gallagher2021generalized}, which shows the differences in interpretable horizontal bars that compare two texts. Particularly, we use a Frequency-based proportion shift method, that consists on measuring the difference of relative frequencies of a word in each text. In Figures \ref{fig:ws1}, \ref{fig:ws2} and \ref{fig:ws3} we show such wordshift graphs for each peak.

\begin{figure}[!htbp]
    \centering
    \includegraphics[width=0.45\linewidth]{figures_SI/peakA.png}
    \includegraphics[width=0.45\linewidth]{figures_SI/peakB.png}
    \caption{\textbf{Wordshift graphs for peak A (left) and peak B (right).} }
    \label{fig:ws1}
\end{figure}

\begin{figure}[!htbp]
    \centering
    \includegraphics[width=0.45\linewidth]{figures_SI/peakCDE.png}
    \includegraphics[width=0.45\linewidth]{figures_SI/peakF.png}
    \caption{\textbf{Wordshift graphs for peak CDE (left) and peak G (right).}}
    \label{fig:ws2}
\end{figure}

\begin{figure}[!htbp]
    \centering
    \includegraphics[width=0.45\linewidth]{figures_SI/peakG.png}
    \caption{\textbf{Wordshift graph for peak F.}}
    \label{fig:ws3}
\end{figure}

\bibliography{sn-bibliography}